# The SINS survey of z~2 galaxy kinematics: properties of the giant star forming clumps[1]

Genzel, R. [1,2], Newman, S.[3], Jones, T.[3], Förster Schreiber, N.M.[1], Shapiro, K.[3,14], Genel, S.[1], Lilly, S.J.[4], Renzini, A.[5], Tacconi, L.J.[1], Bouché, N.[6,15], Burkert, A.[7], Cresci, G.[8], Buschkamp, P.[1], Carollo, C.M.[4], Ceverino,D.[9], Davies, R.[1], Dekel,A.[9], Eisenhauer, F.[1], Hicks, E.[10], Kurk, J.[1], Lutz, D.[1], Mancini, C.[5], Naab, T.[11], Peng, Y.[4], Sternberg, A.[12], Vergani, D.[13] & Zamorani, G.[13]

[1] *Max-Planck-Institut für extraterrestrische Physik (MPE), Giessenbachstr.1, D-85748 Garching, Germany (genzel@mpe.mpg.de)*

[2] *Department of Physics, Le Conte Hall, University of California, Berkeley, CA 94720, USA*

[3] *Department of Astronomy, Campbell Hall, University of California, Berkeley, CA 94720, USA*

[4] *Institute of Astronomy, Department of Physics, Eidgenössische Technische Hochschule, ETH Zürich, CH-8093, Switzerland*

[5] *Osservatorio Astronomico di Padova, Vicolo dell'Osservatorio 5, Padova, I-35122, Italy*

[6] *Department of Physics & Astronomy, University of California, Santa Barbara, Santa Barbara, CA 93106, USA*

[7] *Universitäts-Sternwarte Ludwig-Maximilians-Universität (USM),* Scheinerstr. 1, München, D-81679, Germany

[8] *Istituto Nazionale di Astrofisica – Osservatorio Astronomico di Arcetri, Largo Enrico Fermi 5, I - 50125 Firenze, Italia*

[9] *Racah Institute of Physics, The Hebrew University, Jerusalem 91904, Israel*

[10] *Department of Astronomy, University of Washington, Box 351580, U.W., Seattle, WA 98195-1580, USA*

[11] *Max-Planck Institute for Astrophysics, Karl Schwarzschildstrasse 1, D-85748 Garching, Germany*

[12] *School of Physics and Astronomy, Tel Aviv University, Tel Aviv 69978, Israel*

[1] Based on observations at the Very Large Telescope (VLT) of the European Southern Observatory (ESO), Paranal, Chile (ESO program IDs 076.A-0527, 079.A-0341, 080.A-0330, 080.A-0339, 080.A-0635, 183.A-0781).

[13] *INAF Osservatorio Astronomico di Bologna, Via Ranzani 1, 40127 Bologna, Italy*

[14] *Aerospace Research Laboratories, Northrop Grumman Aerospace Systems, Redondo Beach, CA 90278, USA*

[15] *supported by the Marie Curie grant PIOF-GA-2009-236012 from the European Commission*

# Abstract

We have studied the properties of giant star forming clumps in five z~2 star-forming disks with deep SINFONI AO spectroscopy at the ESO VLT[1]. The clumps reside in disk regions where the Toomre Q-parameter is below unity, consistent with their being bound and having formed from gravitational instability. Broad Hα/[NII] line wings demonstrate that the clumps are launching sites of powerful outflows. The inferred outflow rates are comparable to or exceed the star formation rates, in one case by a factor of eight. Typical clumps may lose a fraction of their original gas by feedback in a few hundred million years, allowing them to migrate into the center. The most active clumps may lose much of their mass and disrupt in the disk. The clumps leave a modest imprint on the gas kinematics. Velocity gradients across the clumps are 10-40 km/s/kpc, similar to the galactic rotation gradients. Given beam smearing and clump sizes, these gradients may be consistent with significant rotational support in typical clumps. Extreme clumps may not be rotationally supported; either they are not virialized, or they are predominantly pressure supported. The velocity dispersion is spatially rather constant and increases only weakly with star formation surface density. The large velocity dispersions may be driven by the release of gravitational energy, either at the outer disk/accreting streams interface, and/or by the clump migration within the disk. Spatial variations in the inferred gas phase oxygen abundance are broadly consistent with inside-out growing disks, and/or with inward migration of the clumps.

.

# 1. Introduction

The rest-frame UV/optical morphologies of most z > 1 'normal' star forming galaxies (henceforth '*SFGs*': Steidel et al. 1996, 2004, Franx et al. 2003, Noeske et al. 2007, Daddi et al. 2007, Cameron et al. 2010) are irregular and often dominated by several giant (kpc-size) star forming clumps (Cowie et al. 1995, van den Bergh et al. 1996, Elmegreen et al. 2004, 2009, Elmegreen & Elmegreen 2005, 2006, Förster Schreiber et al. 2009, 2011a). These clumpy, asymmetric structures often resemble z~0 mergers (Conselice et al. 2003, Lotz et al. 2004). However, spatially resolved studies of the ionized gas kinematics of these clumpy galaxies find a surprisingly large abundance of disks with coherent rotation, especially among the more massive ($M_* \geq$ a few $10^{10}$ $M_\odot$) and bright ($K_{s\,AB} \leq 21.8$) systems (Förster Schreiber et al. 2006, 2009, Genzel et al. 2006, 2008, Weiner et al. 2006, Wright et al, 2007, Law et al. 2007, 2009, Shapiro et al. 2008, Bournaud et al. 2008, Cresci et al. 2009, van Starkenburg et al. 2008, Epinat et al. 2009, Lemoine-Busserolle & Lamareille 2010). These kinematic studies also find that high-z SFGs as a rule exhibit large local velocity dispersions of their ionized gas component, with ratios of rotation velocity $v_c$ to local intrinsic velocity dispersion $\sigma_0$ ranging from 1 to 6. Observations of CO rotational line emission indicate that z~1-3 SFGs have large (~30-80%) baryonic cold gas fractions (Daddi et al. 2008, 2010a, Tacconi et al. 2008, 2010).

These basic observational properties can be understood in a simple physical framework, in which gravitational instability and fragmentation in semi-continuously fed, gas-rich disks naturally leads to large turbulence and giant star forming clumps (Noguchi 1999, Immeli et al. 2004 a,b, Bournaud, Elmegreen & Elmegreen 2007,

Elmegreen et al. 2008, Genzel et al. 2008, Dekel, Sari & Ceverino 2009, Bournaud 2010). A more detailed discussion of these instabilities follows in section 2.4 where we show that gas rich, marginally stable disks should have much larger and more massive star forming complexes than those in z~0 SFGs and that these complexes should be located in regions where the value of the Toomre (1964) *Q*-parameter is below unity.

The most recent generation of cosmological galaxy evolution models and simulations find that the buildup of z>1 SFGs in the mass range of $10^{10}$ to $10^{11}$ $M_\odot$ is dominated by smooth accretion of gas and/or minor mergers (Kereš et al. 2005, 2009, Dekel & Birnboim 2006, Bower et al. 2006, Kitzbichler & White 2007, Ocvirk, Pichon & Teyssier 2008, Davé 2008, Dekel et al. 2009a, Oser et al. 2010). In contrast the overall cosmological mass assembly of galaxies, especially of the most massive ones and at late times, is probably dominated by mergers (Bower et al. 2006, Kitzbichler & White 2007, Naab et al. 2007, Naab, Johansson & Ostriker 2009, Guo & White 2008, Davé 2008, Genel et al. 2008). The large and semi-continuous gas accretion in these 'cold flows' or 'cold streams' may rapidly build up galaxy disks (Dekel et al. 2009a, Ocvirk et al. 2008, Kereš et al. 2009, Oser et al. 2010). If the incoming material is gas rich, then violent gravitational instabilities in these disks could lead to the large star formation rates derived from observations (Genel et al. 2008, Dekel et al. 2009b). The giant clumps are expected to migrate into the center via dynamical friction and tidal torques on a time scale of

$$t_{inspiral} \approx \left( {}^{v_c}\!/\!_{\sigma_0} \right)^2 t_{dyn}(R_{disk}) \sim 10\ t_{dyn}(R_{disk}) < 0.5\ \mathrm{Gyr} \qquad (1),$$

where they may form a central bulge and a remnant thick disk (Noguchi 1999, Immeli et al. 2004 a,b, Förster Schreiber et al. 2006, Genzel et al. 2006, 2008, Elmegreen et al. 2008, Carollo et al. 2007, Dekel et al. 2009b, Bournaud, Elmegreen & Martig 2009, Ceverino, Dekel & Bournaud 2010).

The efficacy of the 'violent disk instability' for forming bulges by the in-spiral of the giant clumps hinges on the survival of the clumps in the presence of outflows driven by stellar winds, supernovae and radiation pressure, even if secular bulge growth may also occur directly from the disk without clump migration. This 'star formation feedback' is widely thought to be a key ingredient in the evolution of star forming galaxies (Dekel & Silk 1986, Kauffmann, White & Guiderdoni 1993, Finlator & Davé 2007, Efstathiou 2000, Bouchè et al. 2010, Dutton, van den Bosch & Dekel 2010). Local Universe giant molecular clouds (GMCs) are prone to rapid expulsion of gas by this feedback on a time scale $t_{\rm expulsion} \sim M_{clump} / \dot{M}_{out}$, which probably dissipates GMCs on a time scale of a few tens of Myrs (Murray 2010). High-z clumps may live longer because their ratio of gravitational binding energy to star formation rate is ~100 times larger than in the local Universe (Dekel et al. 2009b). Exactly how stable the high-z clumps are and how large their gas expulsion time scales might be, is a matter of current debate. Krumholz & Dekel (2010) find that the high-z clumps only loose a modest fraction (<50%) of their original mass by feedback as long as the star formation efficiency per free fall time does not significantly exceed a few percent (similar to local SFGs: Kennicutt 1998a). Murray, Quataert & Thompson (2010a) and Genel et al. (2010) argue that the majority of the clumps' initial gas mass is expelled by feedback in the form of momentum driven winds.

While it is plausible that the very active high-z SFGs are naturally driven toward marginal gravitational instability ($Q$~1) by self-regulation (Quirk 1972, Gammie 2001, Thompson, Quataert & Murray 2005), the dominant agents responsible for the required (and observed) high velocity dispersions are not known, and possibly multi-factorial (Krumholz & Burkert 2010). Förster Schreiber et al. (2006) proposed that the gravitational energy released by the accreting gas (including minor mergers) at the interface of the cold streams and the disk may trigger the large random motions. A similar explanation is favored by Genzel et al. (2008) and Khochfar & Silk (2009), while Dekel et al. (2009b) argue that smoother-than-average streams may not be able to drive a large local velocity dispersion but in other cases accretion from the halo might drive the disk into stability (Q>1). Instead, Immeli et al. (2004 a,b), Dekel et al. (2009b) and Ceverino et al. (2010) all favor gravitational torques in the disk and collisions between the giant clumps, or a combination of the gravitational torques and stellar feedback (Elmegreen & Burkert 2010) as the main drivers of the turbulence. If the main driver of the large velocity dispersions is stellar feedback, and specifically radiation pressure on dust grains, one might expect a correlation of the amplitude of turbulence with star formation rate or surface density (Förster Schreiber et al. 2006, Genzel et al. 2008, Murray et al. 2010a).

In this paper we present and analyze new high-quality SINFONI/VLT integral field (IFU) spectroscopy (Eisenhauer et al. 2003, Bonnet et al. 2004) of five luminous, clumpy z~2 SFGs. We employed both laser guide star (LGS) and natural guide star (NGS) adaptive optics (AO) to improve the angular resolution to an effective ~0.2" FWHM. For all of the targets, the quality of the derived spectra is much superior to previous data, because of long integration times (9 to 19 hours in four of the targets) and/or the high surface brightness of the selected clumpy galaxies. With these data it is now possible, for the first time, to study detailed line profiles on the scale of the most massive and largest clumps (1-3 kpc). Our measurements deliver interesting new constraints on the kinematic properties and lifetimes of the giant clumps. We adopt a ΛCDM cosmology with $\Omega_m$=0.27, $\Omega_b$=0.046 and $H_0$=70 km/s/Mpc (Komatsu et al. 2010), as well as a Chabrier (2003) initial stellar mass function (IMF).

# 2. Observations and Analysis

## *2.1 Source Selection, Observations and Data Reduction*

By selection, the five galaxies we discuss in this paper are massive ($v_c$~250 km/s, $M_*$~$10^{10.6}$ $M_\odot$, $M_{dyn}(R$~10 kpc$)$ ≥$10^{11}$ $M_\odot$), luminous (star formation rates (*SFR)* ~ 120- 290 $M_\odot yr^{-1}$) and fairly large ($R_{disk}$(HWHM)~3-6.5 kpc). They sample the upper range of mass and bolometric luminosity of the z~2 SFG 'main sequence' (Förster Schreiber et al. 2009, Noeske et al. 2007, Daddi et al. 2007). In this subset of the z~2 SFG population, data cubes with integration times a few to twenty hours per galaxy have sufficient signal to noise ratio (SNR) in a sufficient number of independent pixels ($N_{pix}$~50-200) to extract the detailed properties of giant clumps, each of which have intrinsic FWHM diameters of 0.15"-0.3" (Genzel et al. 2008).

As part of the SINS GTO survey (Förster Schreiber et al. 2009) and the SINS/zCOSMOS ESO Large Program (see Mancini et al., in prep.) of high-z galaxy kinematics carried out with SINFONI at the VLT, we observed the Hα, [NII] and [SII] emission lines in the rest-frame UV-selected SFGs Q1623-BX599 (z=2.332) and Q2346-BX482 (z=2.258: Erb et al. 2006b, Förster Schreiber et al. 2006, 2009), and in the rest-frame optically selected SFGs D3a15504 (z=2.383) ; ZC782941 (z=2.182) and ZC406690 (z=2.196) (Kong et al. 2006; Genzel et al. 2006, 2008; Förster Schreiber et al. 2009; Mancini et al., in prep; Peng et al., in prep.). The two rest-UV-selected sources were photometrically identified in optical imaging by their $U_nGR$ colors (satisfying the 'BX' criteria), their redshift confirmed from optical spectroscopy, and first observed in the near-IR with the long-slit spectrometer NIRSPEC on Keck II (Steidel et al. 2004; Adelberger et al. 2004; Erb et al. 2006b). The rest-optically-selected targets were identified based on $K_s$-band imaging and via the 'BzK' color criterion for 1.4 < z < 2.5 star-forming galaxies (Daddi et al. 2004b), and followed-up with VLT/VIMOS optical spectroscopy to confirm their redshift (Kong et al. 2006; Lilly et al. 2007). Prior to the SINFONI observations, none of them had near-IR spectroscopic data. ZC782941 and ZC406690 were moreover specifically drawn from the 1.7 $deg^2$ zCOSMOS spectroscopic survey (Lilly et al, 2007) to be located within 30" of $G < 16$ mag stars suitable for Natural Guide Star adaptive optics (AO) assisted observations.

The five galaxies span the range of kinematic properties found in the SINS survey of z~2 SFGs (Förster Schreiber et al. 2009). BX482 and ZC406690 are large clumpy, rotating disks with a prominent ~5kpc ring of star formation. D3a15504 is a large rotating disk with a central AGN. ZC782941 is a more compact, rotating and asymmetric disk. The asymmetry is mainly caused by a compact clump north of the main body the galaxy, which may be a second, lower mass galaxy interacting with the main galaxy (a 'minor' merger). BX599 is an example of the compact 'dispersion dominated' systems that tend to be common among less massive, UV-selected galaxies (Erb et al. 2006b, Law et al. 2007, 2009). However, our new LGS AO SINFONI data now resolve BX599 spatially and reveal a substantial velocity gradient of 150 to 200 km/s across ~3 kpc. The observed ratio of half the velocity gradient to the integrated velocity dispersion $\Delta v_{grad}/(2\sigma_{int})$~0.6. This is similar to several rotating disk galaxies in the SINS survey (Förster Schreiber et al. 2009). BX599 may thus be a compact rotating disk. For a more detailed description of the SINS and SINS/zCOSMOS surveys, source selection, and global galaxy properties, we refer to Förster Schreiber et al. (2009) and Mancini et al. (in prep.).

Table 1 summarizes integration times and the final FWHM angular resolutions in these galaxies. For a description of the data reduction methods and analysis tools we refer to Schreiber et al. (2004), Davies (2007), and Förster Schreiber et al. (2009). With the final data cubes in hand, we median-filtered the data by two spatial pixels and fitted Gaussian line profiles to each pixel with the fitting code LINEFIT (Förster Schreiber et al. 2009). LINEFIT performs weighted fits to the observed line profiles as a function of the two spatial coordinates based on an input noise data cube and an input spectral response function. The instrumental spectral response function as obtained from OH sky lines is shown as a grey dashed curve in the profiles shown in Figures 7 to 9. For the 0.05"x 0.125" pixel scale in K-band we used here it is fit quite well by a Gaussian of FWHM ~85 km/s (green curve in the upper left panel of Figure 9), with some excess emission in the line wings relative to this best fitting Gaussian. For the analysis of the line profiles in our program galaxies these small differences are negligible, however. LINEFIT takes this instrumental line profile as inputs to compute *intrinsic* velocity dispersions. Likewise the velocity dispersions listed in Table 2 and shown in Figures A1 and A2 are intrinsic values after removal of the instrumental broadening (and any beam smeared rotation).

Uncertainties of all fitted parameters are calculated through 100 Monte-Carlo simulations in which the spectrum of each spatial pixel is perturbed assuming a Gaussian distribution of the rms from the input noise cube. The final integrated line intensity, velocity and velocity dispersion maps were then multiplied by a mask constructed from all pixels with Hα line emission at >3σ significance. We compare the line emission maps to similar resolution (~0.15"-0.25" FWHM) images of the rest-frame UV/optical stellar continuum. In the case of BX482 we use the HST/NIC2-H-band image (through the F160W filter) of Förster Schreiber et al. (2010). For ZC782941 and ZC406690 we use the HST/ACS I-band (F814W filter) images taken as part of the COSMOS survey (Koekemoer et al. 2007). For D3a15504 we have taken and analyzed a 2h AO-assisted exposure of the galaxy with VLT/NACO in $K_s$-band, as part of our original SINS survey program (Förster Schreiber et al. 2009).

### 2.2 Modeling of the velocity fields

We identified the most prominent clumps from maps of individual Hα velocities ('channels') or, in the case of clump D in BX482, from the rest-frame optical continuum map. For identification as a clump, we required the presence of an obvious local maximum in at least two separate velocity channel maps. Figure 1 gives examples of such velocity channel maps for D3a15504 (top row), BX482 (middle rows) and ZC782941 (bottom row), and marks the positions of the most prominent clumps by circles/ovals and alphabetical symbols. Our list of clumps is meant to identify the brightest obvious clumps, and is not complete for the fainter clumps whose identification can be more ambiguous. For BX482 (Figure 2 middle left column), ZC782941 (middle right column), ZC406690 (Figure 2 right column) and BX599 (Figure 9) the brightest clumps also stand out in the velocity integrated Hα and the continuum maps. In D3a15504 and ZC782941 (clumps B-E) some of the clumps are less obvious or even washed out in the integrated maps because of diffuse integrated disk emission. We determined intrinsic HWHM clump radii from Gaussian fits to the appropriate velocity channels and subtracted the instrumental resolution in squares.

In addition to the basic velocity and velocity dispersion maps obtained from LINEFIT, we also constructed 'residual' maps by removing the large scale velocity field. For this purpose we used 'kinemetry' (Krajnović et al. 2006, Shapiro et al. 2008), or simple rotating disk models fitted to the Hα data (Genzel et al. 2006, 2008, Cresci et al. 2009). The resulting velocity/dispersion maps capture the large scale kinematics, which can then be subtracted from the LINEFIT maps, in order to make local residuals stand out more clearly. For the purposes of the analysis presented below, both methods give indistinguishable results.

To perform a kinemetry analysis, we require knowledge of the dynamical center, position angle, and inclination of a galaxy. For the high SNR data presented here, we are able to determine the dynamical centers directly from the shapes of the iso-velocity contours. Position angles and inclinations are estimated from the orientations of the maximum velocity gradients (line of nodes) and the minor to major axis ratios of the line and continuum emission. Using these inputs, we parameterize the observed velocity fields as Fourier expansions along the angle $\varphi$ in the plane of the sky. Ideal, thin-disk rotation is described by a $cos(\varphi)$ term (see Shapiro et al. 2008 for more details). To determine the higher order (local) variations of the velocity field, and/or larger scale, non-axisymmetric deviations from simple rotational motion, we subtract this $cos(\varphi)$ map from the observed velocity field.

The disk models compute data cubes from input structural parameters (c.f. Cresci et al. 2009). For BX482 and ZC406690 we use input models with a ring surrounding a central (extincted) bulge for the mass distribution, and for D3a15504 and ZC782941 we use exponential disk models (Genzel et al. 2006, 2008, Newman et al. 2011, in preparation). Dynamical modeling and analysis of the rest-frame optical morphology indicates that this central component in BX482 has ~20% of the total disk mass (Genzel et al. 2008; Förster Schreiber et al. 2011a, 2011b in preparation). In either case the absence and/or weakness of emission from the center has no influence on the analysis we discuss in the following. Position angles and inclinations are determined as above. The model data are then convolved with the angular and spectral resolution profiles and sampled at the observed pixel scales. The total dynamical mass $M_{dyn}$ is then varied to achieve a best fit match to the observed rotation velocities. To study the non-axisymmetric motions in a galaxy, the best-fit model velocity and velocity dispersion maps are subtracted from the respective observed maps.

We compare these residual maps with Hα surface brightness maps derived from the observed data cubes. Likewise we constructed [NII]/Hα ratio maps from integrated line emission maps smoothed with a 3pixel (0.15”) kernel. We multiplied these maps with a mask constructed from all pixels with Hα emission at >3σ significance. We also constructed pixel-pixel correlation plots of residual velocity dispersion ($\delta\sigma=\sigma$(data)-$\sigma$(model))) vs. Hα surface brightness, and [NII]/Hα line ratio vs. Hα surface brightness. Before investigating possible trends in these correlations, we culled pixels with large δσ or [NII]/Hα uncertainties. Additionally, in the case of D3a15504 (which has a prominent central bulge, AGN and a narrow line region) we also removed the nuclear region.

## 2.3 Determination of star formation rates and gas masses

For calculating star formation rates and gas surface densities from the Hα data we used the conversion of Kennicutt (1998b) modified for a Chabrier (2003) IMF (*SFR*=$L(H\alpha)_0/2.1\times10^{41}$ erg/s). We corrected the observed Hα fluxes for spatially uniform extinction with a Calzetti (2001) extinction curve (A(Hα)=7.4 E(B-V)), including the extra ‘nebular’ correction ($A_{gas}=A_{stars}/0.44$) introduced by Calzetti (2001). We determined E(B-V) from the integrated UV/optical photometry of the galaxies (row 5 in Table 2). Förster Schreiber et al. (2009) find that including the extra nebular correction brings Hα- and UV-continuum based star formation rates of z~2 SINS galaxies into best agreement.

We estimated molecular surface densities (and masses, including a 36% helium contribution) from equation (8) of Kennicutt et al. (2007), modified for the Chabrier IMF used here,

$$\log\left(\frac{\Sigma_{mol-gas}}{M_{\odot}pc^{-2}}\right)=0.73\ \log\left(\frac{\Sigma_{star-form}}{M_{\odot}yr^{-1}kpc^{-2}}\right)\ +2.91 \qquad (2).$$

Equation (2) is based on Hα, 24μm and CO observations of M51 and is similar to results for larger samples of z~0 SFGs (e.g. equation (4) in Kennicutt 1998a, and Figure 4 of Genzel et al. 2010). It has the added advantage of being based on spatially resolved measurements of the gas to star formation relation with a similar spatial resolution (0.5 kpc) as our high-z data and also covering a similar range of gas surface densities (10-$10^3$ $M_{\odot}pc^{-2}$). Figure 4 in Genzel et al. (2010) (see also Daddi et al. 2010b) also shows that to within the uncertainties (of about a factor of 2), z~0 and z~1-3 SFGs (with galaxy integrated measurements of CO luminosities and *SFR*s) are fit by the same relation, although the gas masses from the best fits of Genzel et al. (2010) are ~20% larger than estimated from equation (2). In equation (2) we did not correct the data for the fraction of Hα emission from outflowing gas (see section 3.2 below). This correction is small, with the exception of the brightest clumps where gas surface densities may be somewhat overestimated.

The gas surface densities/masses and star formation rates estimated from equation (2) and in listed Table 2 are uncertain by at least a factor of 2 to 3. In addition to the well known issue of how to infer molecular gas column densities/masses from the integrated line flux of an optically thick CO rotational line (see the in depth discussion in Tacconi et al. 2008, Genzel et al. 2010), and the question of whether equation (2) adequately describes the gas to star formation relation for the physical conditions on clump scales at z~2, there is the important issue of differential extinction. We will argue in section 3.2 that the asymmetry of broad Hα/[NII] line emission is direct evidence for such differential extinction. It is unclear, however, what the general impact of the differential extinction would be on clump scales. One might naively expect that the effect increases gas column densities/masses relative to averages on larger scales. However, there are almost certainly also evolutionary effects, such that in a given aperture there may both be very high dust column densities in neutral clouds, as well as HII regions with relatively low extinction. Such spatial separations of 300 pc to >1 kpc are seen in nearby spirals, such as M51 (Rich & Kulkarni 1990), as well as at z~1 (Tacconi et al. 2010). As a result, the Kennicutt-Schmidt scaling relation in equation (2) may break down or be significantly altered on small scales (e.g. Schruba et al. 2010 in M33 on ≤ 80 pc scales).

## *2.4 Spatial distribution of the Toomre Q-parameter*

A rotating, symmetric and thin gas disk is unstable to gravitational fragmentation if the Toomre *Q*-parameter (Toomre 1964) is ≤1. For a gas dominated disk in a background potential (of dark matter and an old stellar component) *Q* is related to the local gas velocity dispersion $\sigma_0$ (assuming isotropy), circular velocity $v_c$, epicyclic frequency $\kappa$ ($\kappa^2 = 4(v_c/R_{disk})^2 + R_{disk}\, d(v_c/R_{disk})^2/dR_{disk}$), gas surface density $\Sigma_{gas}$, and radius of the disk $R_{disk}$ via the relation (Binney & Tremaine 2008, Escala & Larson 2008, Elmegreen 2009, Dekel et al. 2009b)

$$Q_{gas} = \frac{\sigma_0 \kappa}{\pi G \Sigma_{gas}} = \left(\frac{\sigma_0}{v_c}\right)\left(\frac{a\left(v_c^2 R_{disk}/G\right)}{\pi R_{disk}^2 \Sigma_{gas}}\right) = \left(\frac{\sigma_0}{v_c}\right)\left(\frac{aM_{tot}}{M_{gas}}\right) = \left(\frac{\sigma_0}{v_c}\right)\left(\frac{a}{f_{gas}}\right) \qquad (3).$$

Here the constant $a$ takes on the value of 1, $\sqrt{2}$, $\sqrt{3}$ and 2 for a Keplerian, constant rotation velocity, uniform density and solid body disk; $f_{gas}$ is the gas fraction within $R_{disk}$. If the disk consists of molecular ($H_2$+He), atomic (HI+He) and stellar (*) components, $Q_{tot}^{-1} = Q_{H2}^{-1} + Q_{HI}^{-1} + Q_*^{-1}$ if all components have similar velocity dispersions. If there is a (young) stellar component distributed similarly to the gas, the combined gas + young star component will thus have a $Q_{tot}$ that is inversely proportional to the sum of the gas and stellar surface densities. In that case $f_{gas}$ should be replaced by the mass fraction $f_{young}$ of that 'young' component. Such a disk is unstable (or stable) to fragmentation by gravity depending on whether $Q_{tot}$ is less (or greater) than unity. Equation (3) can be rewritten as

$$\left(\frac{\sigma_0}{v_c}\right) = \left(\frac{z}{R_{disk}}\right) = \frac{Q\, f_{young}}{a} \qquad (4),$$

where $z$ is the z-scale height of the disk. Gas rich, marginally stable disks are thick and turbulent. The largest and fastest growing, Jeans-unstable mode not stabilized by rotation is the 'Toomre scale/mass', given by (Elmegreen 2009, Genzel et al. 2008, Escala & Larson 2008, Dekel et al. 2009b)

$$R_{Toomre} \approx 0.8\, Q^{-1} a^{-2} \left(\frac{\sigma_0}{v_c}\right) R_{disk} \approx 1 \left(\frac{f_{young}}{0.4}\right)\left(\frac{R_{disk}}{5\,kpc}\right) \text{kpc} \propto \frac{\sigma_0^2}{\Sigma_{gas}} \quad \text{and}$$
$$M_{Toomre} \approx 0.6\, Q^{-2} a^{-4} \left(\frac{\sigma_0}{v_c}\right)^2 M_{disk} \approx 5x10^9 \left(\frac{f_{young}}{0.4}\right)^2 \left(\frac{M_{disk}}{10^{11}\, M_\odot}\right) M_\odot \propto \frac{\sigma_0^4}{\Sigma_{gas}} \quad (5),$$

where the numerical factors are for a flat rotation curve ($a$=1.4). Gas rich, marginally stable disks thus should have much larger and more massive star forming complexes than those in z~0 SFGs with (cold) gas fractions of less than 10% and larger fractions of stabilizing old stellar disks and bulges.

For the four well resolved disks/rings, we created maps of the Toomre parameter *Q(x,y)*. We combined the computed gas surface density for each pixel (equation 2), with the best fitting model rotation curve to compute the epicyclic frequency $\kappa$ and the velocity dispersion map to calculate *Q(x,y)* from equation (3). We then used different Monte-Carlo realizations and standard error propagations to compute maps of the uncertainties $\Delta Q$.

# 3. Results

Figure 1 and Figure 2 show velocity channel maps and the integrated Hα and continuum images for four of the five galaxies. The integrated Hα image of the fifth galaxy (BX599) is shown in the top center panel of Figure 8. In this case, we do not have access to a high resolution continuum image. The most prominent clumps are labeled for each galaxy (see the more detailed discussion in section 2.2). Tables 2 and 3 summarize the derived physical properties. A 'typical' individual clump within the massive ($M_*\sim10^{10....11}$ $M_\odot$) BX/BzK galaxies in the SINS survey, such as an average clump in D3a15504, ZC782941 and BX482, accounts for a few percent of the UV/optical light of the entire galaxy, has a current star formation rate of a few solar masses per year, and a stellar mass of one to a few times $10^9$ $M_\odot$ (Table 2, Förster Schreiber et al. 2011b, in prep.). The most extreme clumps in BX482 and ZC406690 make up ~10-20% of the integrated Hα fluxes, have star formation rates of 10 to 40 solar masses per years and masses $\sim10^{10}$ $M_\odot$.

## 3.1 Giant clumps are the locations of gravitational instability

As discussed in the Introduction and section 2.4, a plausible hypothesis is that the ~1-2 kpc diameter giant star forming clumps in z>1 SFGs represent the largest/most massive gravitationally unstable entities in the high-z disks. If this is indeed the case an empirical determination of the Toomre parameter (equation 2) as a function of position should show that clumps and their surroundings have $Q \leq 1$.

Following the methods discussed in the last section, Figures 3 to 6 give the $Q$-maps at a resolution of ~0.22 to 0.25" FWHM for D3a15504, BX482, ZC782941 and ZC406690, where we have only retained pixels with an rms uncertainty $\Delta Q$<0.3 to 0.5. As inputs for our calculations we used the velocity, velocity dispersion and Hα integrated flux maps shown in the left and middle panels of Figures 3 to 6. The central regions in all four galaxies should be neglected, for the following reasons. The central few kpc of D3a15504 may be affected by a central AGN, as well as by large non-circular motions. Both increase the velocity dispersion there (Figure 3 bottom left, Genzel et al. 2006). The central regions of BX482, ZC782941 and ZC406690 exhibit elevated velocity dispersions due to an additional central mass (without much Hα emission) in the cases of BX482 (Genzel et al. 2008) and ZC406690 (Newman et al. 2011, in preparation), and unresolved beam smearing of rotation in ZC782941.

We find that throughout the extended outer disks and toward the clumps of D3a15504, BX482, ZC782941 and ZC406690 the empirically determined $Q$-parameter is at or even significantly below unity. As postulated, these SFGs are indeed unstable to fragmentation throughout their disks. The clumps are thus gravitationally bound or nearly so. Our analysis only considers the gaseous component. As discussed in 2.4, taking into account a stellar component with dispersion similar to that of the gas will probably lower the $Q$-values still further. Given the typical molecular gas fractions of ~0.3 to 0.8 (Tacconi et al. 2010, Daddi et al. 2010), this pushes $Q$ to significantly below unity in the prominent clumps. These clumps thus appear to be in the highly unstable regime, where linear Toomre-stability analysis is inappropriate. The fact that the $Q$-parameter is below unity even in the more diffuse disk regions suggest that global perturbations are significant in setting the $Q$-distribution. We conclude that the $Q$-maps in Figures 3 to 6 are consistent with the commonly held view that the clumps form by gravitational instability. However, we cannot exclude the alternative possibility that the instability is driven by a large scale compression, such as experienced in a galaxy interaction or (minor) merger (e.g. Di Matteo et al. 2007).

## 3.2 Evidence for powerful outflows on clump scales

UV spectroscopy of metal absorption lines and of Lyα emission lines provide compelling evidence for ubiquitous mass outflows in ‘normal’ high-z (Pettini et al. 2000, Shapley et al. 2003, Steidel et al. 2004, 2010, Weiner et al. 2009). More recently, Shapiro et al. (2009) have reported broad Hα/[NII] optical line emission from stacking of 47 z~2 SFG spectra in the SINS survey. In both cases the information is integrated over galaxy (or intergalactic) scales and, in the case of absorption lines, the location of the absorber along the line of sight is not or only approximately (Steidel et al. 2010) constrained. Lyα is strongly self-absorbed so that any information on spatial distribution and kinematics is washed out and strongly depends on modeling. The broad lines discussed by Shapiro et al. (2009) could originate in extended galactic winds coming from star forming disks (similar to z~0 starburst galaxies: Armus, Heckman & Miley 1990, Lehnert & Heckman 1996, Martin 1999, 2005, Rupke, Veilleux & Sanders 2005, Veilleux , Cecil & Bland-Hawthorne 2005, Strickland & Heckman 2009) or alternatively, in outflows driven by a central AGN.

### 3.2.1 broad wings of Hα emission associated with clumps

Our data provide for the first time direct evidence for powerful outflows on the scale of individual star forming clumps. Figures 7 and 8 show extracted spectral profiles for individual bright clumps, as well as for the entire galaxy in Q1623-BX599 and ZC400690. We find that

- the prominent clumps A and B in ZC406690 exhibit blue line wings extending 500 and 1000 km/s half width at zero power (HWZP) from line center;
- a more symmetric, broad component (FWHM 1000 km/s, HWZP ~ 1000 km/s) is seen in the integrated spectrum of ZC406690 and BX599. In the latter the broad emission originates in a compact region (intrinsic diameter ≤ 3kpc) centered within ~1kpc of the peak of narrow Hα emission (clump 'A': top panels in Figure 8);
- somewhat blue-shifted, broad components are also present in clump A of ZC782941, the central pixels of clump A of BX482 and perhaps in the average clump spectrum of D3a15504;
- there are no detectable broad wings in the off-clump galaxy emission of D3a15504 and BX482, or in the southern clumps of ZC782941.

Applying two or multi-component component Gaussian fits, we find that toward the clumps of the five SFGs in Table 2 the broad ($\Delta v(FWHM)_{broad}$~300-1000 km/s), modestly blueshifted ($<v_{broad}>-<v_{narrow}>$~ -30..-150 km/s) components constitute <20 to 60% of the total Hα line flux (row 12 of Table 2). The broad component fits are shown as thin continuous red lines in the spectra in Figures 7 to 9. The evidence for these broad wings is also apparent in the co-added spectrum of the clumps in all five galaxies (Figure 9).

The line widths and flux fractions of the broad emission components in the five galaxies are in good agreement with the stacking results of Shapiro et al. (2009). In the bin of the most massive, highest star formation rate galaxies ($M_*>7x10^{10}\ M_\odot$, comparable to our sample) Shapiro et al. (2009) find $FWHM_{broad}$~2200 (+400,-750) km/s and $f_{broad}$=0.31(+0.04,-0.1). For comparison, the average spectrum in Figure 9 shows that the broad component constitutes about 50% of the average emission from the brightest clumps. The contribution for the galaxy wide averages in our sample is about 30%, comparable to Shapiro et al. (2009). This comparison adds further support to our conclusion that the broad emission comes from individual giant star forming clumps in the disks, rather than from the nuclear regions, or the extended disks.

The fits of broad emission components in Figures 7 through 9 assume that the local 'narrow component' line profiles (after removal of large scale motions) are Gaussians. This assumption needs justification. The detection of the highest velocity gas in the wings is obviously independent of the assumption of line profiles in the core of the line. However, the quantitative determinations of the relative fraction of the line flux in the broad component listed in row 12 of Table 2 and of the line profiles of the broad component do depend on the assumption of the Gaussian shape of the narrow component. The upper panels of Figure 9 give the co-added line profile of the regions between bright clumps (and on weaker clumps) in D3a15504, BX482 and ZC782941. The upper left panel clearly shows that this 'interclump' profile can indeed be well fitted by the instrumental line profile broadened by a Gaussian of dispersion ~75 km/s. There may be a small amount of excess emission in both blue and redshifted line wings but this emission is less than 10-15% of the total flux. The upper right panel of Figure 9 shows a comparison of this interclump spectrum with the co-added 'bright clump' spectrum from the bottom left of Figure 9. The bright clump spectrum is well fit by the interclump profile in its core and red wing but clearly shows highly significant blue excess emission. The profile of the broad component (only associated with the bright clumps) can then be inferred, without assumptions on intrinsic line profiles, by subtracting a suitable fraction F of the interclump profile from the bright clump profile such that the broad component has a reasonably smooth profile near low velocities. The dotted red curve in the bottom right panel of Figure 9 is the result for F=0.75 but the result is very similar for F anywhere between 0.6 and 0.9. The average broad line profile has a FWHM of 500 km/s and is blue shifted by -50 km/s. This test confirms the Gaussian fit approach in Figures 7 and 8.

### 3.2.2 broad [NII] emission

In the cases of clump B, the integrated galaxy profile and perhaps also clump A in ZC406690, the SNR is sufficient to see a blue wing in the 6585 Å [NII] line similar to that in Hα. The ratio $\{[NII]/H\alpha\}_{broad}$ in ZC406690 B/integrated is about 0.3. This ratio is larger than in the narrow component of the star forming clumps in ZC406690 ($[NII]/H\alpha_{narrow}$~0.07 to 0.23). ZC782941 (clump A) may exhibit weak broad [NII] lines as well, as does the co-added bright clump spectrum in the bottom panels of Figure 9.

### 3.2.3 outflow velocities are similar to those found from UV-spectra and in local starbursts

The velocity widths of the blue-shifted wings in our sample are comparable to those seen in the UV metal-line absorption components in z~2-3 SFGs studied by Pettini et al. (2000) and Steidel et al. (2010). Steidel et al. (2010) find that the velocity centroids of the outflowing gas range between 0 (systemic) and -500 km/s, with an average (in 89 z~2 BX galaxies) of -160 km/s. The centroid of the blueshifted component in Figure 9 is -80±20 km/s, comparable to the values found by Steidel et al. (2010). The maximum blue-shifted velocities ($\Delta v_{max,1}$ ~$|\langle v\rangle_{broad}$-2*$\sigma_{broad}|$) in our sample range between 380 and 1000 km/s (the average spectrum in Figure 9 has 560 km/s), which are also in good agreement with the $\Delta v_{max}$-*SFR* and $\Delta v_{max}$-$v_c$ relations in z~0 starburst and ultra-luminous galaxies (Martin 2005, Veilleux et al. 2005). For $v_c$~250 km/s Martin (2005) finds in her sample of z~0 ultra-luminous galaxies values of $\Delta v_{max,1}$ between 300 and 750 km/s. Veilleux et al. (2005) and Rupke et al.(2005) define the 'maximum' wind speed as $\Delta v_{max,2}=|\langle v\rangle_{broad}-\Delta v_{broad(FWHM)}/2|$. If this measure is applied to the galaxies in our sample we find values ranging between 260 and 640 km/s (Figure 9 gives 360 km/s), or about 2/3 of the values given in Table 2. For a sample of z~0 luminous and ultra-luminous infrared galaxies Rupke et al. (2005) find $\langle\Delta v_{max,2}\rangle$=300-400 km/s. In all these cases some caution is required since a detailed comparison of column density weighted mean absorption profiles with rms-density weighted emission profiles may be misleading. The $\Delta v_{max}$-*SFR* and $\Delta_{max}$-$v_c$ relations in z~0 starburst and ultra-luminous galaxies have been interpreted as support for a 'cool' outflow driven by the momentum of the intrinsic hot wind due to supernovae and stellar winds (Lehnert & Heckman 1996, Martin 2005, Veilleux et al. 2005), and/or the radiation pressure from the star forming regions onto dust grains mixed with the cold gas (Martin 2005, Murray et al. 2005).

It is thus eminently plausible to conclude that the broad, blueshifted Hα emission in the z~2 SFGs originates in warm ionized outflows triggered by the intense star formation activity in the giant clumps. The alternative interpretation of narrow- or broad-line emission from a central AGN (Shapiro et al. 2009) can be excluded for ZC406690, BX482 and D3a15504. Here the broad emission is associated with individual, off-center clumps participating in the general rotation of the galaxy. An AGN interpretation is possible in ZC782941 and BX599. In ZC782941 clump A is separated from and has a peculiar velocity relative to the main body of the galaxy. This clump may thus be the center of a small intruding galaxy that is interacting with the main part of ZC782941 (i.e. a 'minor' merger). Broad emission from that central spot could conceivably come from a buried AGN. Likewise, in the case of the 'dispersion dominated', compact galaxy BX599 the broad emission comes from the brightest emission component (of two: Figure 8), which may be the center of the overall system.

The prevalence of asymmetric blue-shifted line profiles in Figures 7 to 9 suggests that differential extinction is an important factor in determining line profiles and fluxes. Typical gas column densities in the high-z SFGs range from 700 to 8000 $M_\odot pc^{-2}$ (row 9 in Table 2), corresponding to N(H)~$10^{22.8\text{-}23.9}$ $cm^{-2}$, or visual extinctions of $A_V$~30-400, for dust to gas ratios similar to the Milky Way. Dust opacities at high-z may be smaller because of clumpiness and lower metallicity (Reddy et al. 2010) but are plausibly sufficiently high to account for strong differential extinction of Hα across SFG disks. If this is correct, the interpretation of the blue-shifted emission as originating in an outflow is strengthened further.

### 3.2.4 the outflowing gas is extended on kpc-scales

The case of ZC406690 is particularly favorable for studying the properties of the outflowing component, as the 'disk' emission from the clumps is narrow in velocity (due to the low (30 to $40^0$) inclination) and the surface brightness is high because of the highly clumped emission. Clump A has ~22% of the total Hα emission of the galaxy. In the most spectacular broad-component clump (clump B) the broad emission is clearly offset from the narrow emission component by 0.16” toward the south-east. Toward region D (~0.6” south-east of clump B) the line profile is completely dominated by blue emission. This blue gas cannot be part of the rotating gas ring that characterizes the rest of the narrow Hα emission in ZC400690 (upper left panel of Figure 6). It has a similar center velocity as (but narrower profile than) the broad emission closer to clump B (Figure 7). Region B may thus be dominated by a component in the outflow. The projected broad emission associated with clump B thus is extended over 1.3 to 5 kpc. Given the relatively low inclination of the galaxy, the de-projected extent perpendicular to the galactic plane may be still larger, and may be comparable to the diameter of the star forming ring in ZC406690 (~10 kpc). In contrast the Hα-faint but continuum bright clump C has no discernable broad blue-shifted component (Figure 7). These strong spatial variations of the broad emission, as well as of the equivalent width Hα/2700 Å rest-frame continuum may be due to a combination of variable extinction and/or evolutionary effects in individual clumps.

### 3.2.5 the fraction of broad emission may be correlated with star formation surface density

The incidence of strong broad emission in our sample appears to be most obvious for the highest surface brightness clumps and, in turn, we do not detect the outflow component in the ‘interclump’ spectrum shown in the top left panel of Figure 9. For the 7 clumps in Table 2 and the interclump spectrum there may be correlation between the ratio of broad to narrow Hα component fluxes and star formation surface density. A confirmation (or repudiation) of this tantalizing trend would be interesting. This is because the ratio of broad to narrow components is a measure of the ratio of outflow rate to star formation rate. Simple theoretical arguments for both energy and momentum driven winds lead to the expectation that this ratio should be approximately constant (Heckman 2003, Veilleux et al. 2005, Murray et al. 2005). Outflows tap a fraction of the energy and/or momentum (both proportional to *SFR*) released by the young, massive stars. If the expansion velocity of the ionized gas does not strongly depend on *SFR*, the ratio of outflow rate to *SFR* is approximately constant as well. Martin (2005) and Weiner et al. (2009) find $v_{out} \propto SFR^{0.3}$. The star formation surface density is proportional to gas surface density if the near-linear KS-relation of Genzel et al. (2010) and Daddi et al. (2010b) applies. Gas surface density in turn is proportional to dust surface density in dusty sources, which in turn is proportional to dust opacity. A correlation of the ratio of broad to narrow Hα emission with $\Sigma_{star\,form}$, may support the proposal (Murray et al. 2005, 2010a) that the cold/warm outflows in massive star forming galaxies are driven to a significant extent by radiation pressure ($\dot{M}_{out}\, v_{out} \sim \tau_{dust} L/c$). Murray, Ménard & Thompson (2010b) show that in that case galactic winds can only be launched for star formation surface densities above a critical value, $\Sigma_{\mathrm{star\ form,crit}} > 0.1\, v_{c,250km/s}^{2.5}\, R_{5kpc}^{-2}\ \mathrm{M_{\odot}yr^{-1}kpc^{-2}}$. All five galaxies are above this limit.

Given the emerging evidence for powerful winds from individual clumps, the obvious next question is whether this ‘stellar feedback’ is the key agent driving the large-scale turbulence in high-z galactic disks, as proposed by Ostriker & Shetty (2011) and Lehnert et al. (2009). The next section will show, surprisingly perhaps, that this is not evident from our data.

### *3.3 Is the high velocity dispersion of z>1 SFGs driven by star formation feedback?*

Are the large local rms-velocity dispersions (i.e. local FWHM line widths) in ionized gas, germane to all high-z SFGs studied so far, a direct result of the clump outflows discussed in the last section? The large velocity dispersion may, for instance, be driven by the mixing of the hot wind fluid with cooler clouds at the base of the outflows (Westmoquette et al. 2007). If so, one would naively expect a correlation between the rms-dispersion and the surface brightness of Hα, as a measure of star formation surface density (but see Ostriker & Shetty 2011).

There clearly is a correlation of rms line width and the powerful outflows in the extreme clumps ZC406690 A and B (bottom panel in Figure A1). But this increased line width is entirely due to the broad outflow component while the narrow line emission in these clumps does not vary significantly relative to the surrounding.

Figure 10 shows galaxy wide and clump averages of $\sigma_0$ *(or* $\sigma_{clump}$) as a function of $\Sigma_*$, for those z>1 SFGs with good quality determinations. These include the best z~1.5-2.5 disks from the SINS survey (Cresci et al. 2009, Förster Schreiber et al. 2009), and other recent surveys of z~1-2.5 SFGs (Wright et al. 2007, van Starkenburg et al. 2008, Epinat et al. 2009, Lemoine-Busserolle & Lamareille 2010), mainly sampling fairly massive ($M_{dyn}>10^{10.5}\ M_\odot$) galaxies with radii $R_{1/2}$~ 2-10 kpc. We also include galaxy wide averages of lower mass disks (~ a few $10^9\ M_\odot$) from the survey of z~1-3 lensed SFGs by Jones et al. (2010), and from mostly lower mass ($M_{dyn}$~0.3-3x$10^{10}M_\odot$) and compact ($R_{1/2}$~0.9-2 kpc) but well resolved (with AO) 'dispersion dominated' z~1.5-3 SFGs taken from Law et al. (2009). Finally we include our individual clump measurements in Table 2. This compilation samples a wide range of gravitational potentials and star formation surface densities, from scales somewhat more active than ‘normal’ z~0 SFGs (a few $10^{-2}$ $M_{\odot}yr^{-1}$ $kpc^{-2}$) to the ‘Meurer’ limit (~20 $M_{\odot}$ $yr^{-1}$ $kpc^{-2}$). Above this limit there appear to be no or few UV/optically bright SFGs at any redshift (Meurer et al. 1997), with the exception of compact, gas rich mergers at both low-z (ULIRGs) and high-z (submillimeter galaxies {SMGs}).

The rms-velocity dispersion does appear to increase with star formation surface density but only by a modest amount. The measurements included in Figure 10 sample more than two orders of magnitude in $\Sigma_*$, yet $\sigma_0$ changes by less than a factor of 2. A formal weighted fit only yields a marginally significant positive correlation (log($\sigma_0$) ~ 0.039 (±0.022) $log(\Sigma_*)$). An un-weighted fit to the same data gives a steeper slope (0.07±0.025), as does a fit to only the SINS galaxies and clumps (0.1±0.04), or a fit with only AO data sets from SINS and OSIRIS (0.12±0.04). The scatter at any fixed star formation surface density is almost as large as the overall trend, and is formally larger than the measurement errors. This large scatter is at the root of the marginal significance (< 3 standard deviations) of the overall correlation. It is not clear whether the overall trend of the high-z points connects to the region occupied by lower surface star formation density, z~0 SFGs, as presented by Dib, Bell & Burkert (2006).

Recently, Green et al. (2010) have reported Hα integral field spectroscopy in a sample of lower mass, Hα bright star forming galaxies at z~0.1 (including a number of mergers), at a similar linear resolution (~ 2 kpc) as the z~2 AO data sets in this paper. From their analysis Green et al. infer that the *luminosity weighted*, average velocity dispersions $\sigma_L$ scale with star formation rates and, in their most active systems, take on values similar to those seen in high-z galaxies. They conclude that feedback is the main agent driving galactic turbulence at all redshifts. Unfortunately, it is not possible to directly compare $\sigma_L$ to the local velocity dispersions $\sigma_0$ that we are discussing in this paper. The luminosity weighted quantity $\sigma_L$ places the strongest weight on the bright central regions in each galaxy, where beam smearing in rotating disks creates artificially large velocity dispersions, which is not or only partially removed in the analysis of Green et al. (2010). The quantity $\sigma_L$ thus necessarily is an upper bound to $\sigma_0$. A more detailed comparison of the Green et al. data set with our data is highly desirable but requires the application of the same data analysis methods, which is beyond the scope of this paper (Davies et al. in preparation).

We have also looked for possible pixel-to-pixel variations of $\sigma_0$ in the deep AO-data on the four most extended SFGs reported in this paper. There clearly is an increase of the rms dispersion toward clumps A and B in ZC406690 but this increase appears to be entirely due to the powerful outflow component in these extreme clumps. The width of the narrow component does not vary much in ZC406690. We find weak positive correlations between the residual velocity dispersion and Hα surface brightness in D3a15504, and possibly ZC782941, consistent with the trends in Figure 10. There is no dependence of velocity dispersion on Hα surface brightness in BX482 (Appendix A. Figures A1 and A2). To first order, the large velocity dispersions in high-z SFGs appear to form a spatially constant 'floor'. Any differences in $\sigma_0$ between intra- and inter-clump regions are not or only marginally significant, given the 1σ measurement errors of typically 10 to 20 km/s (Figures A1 and A2).

Finally we have checked for a dependence of $\sigma_0$ on mass outflow rate, discussed in section 4.1 and listed in rows 20, 21 & 22 of Table 2. There is little evidence for such a correlation. As the spectra in Figures 7 to 9 show, there is an order of magnitude range of inferred outflow velocities at roughly constant rms line width for the different clumps in ZC406690, and the brightest clumps in BX482 and ZC782941 have comparable outflow rates but very different line widths.

We would like to emphasize that this relatively weak dependence of $\sigma_0$ on star formation surface density does not constitute an inconsistency with the detection of the broad Hα emission discussed in section 3.2. This is because the broad line wings in Figures 7 to 9 do not greatly affect the FWHM line widths in most of the clumps (with the exception of clumps A and B in ZC406690), which are dominated by the narrow component tracing star formation. Again with the exception of ZC406690 the inter-clump regions are not affected at all. However, for the galaxy wide estimates $\sigma_0$ is by necessity estimated from the linewidths in the off-center parts of the galaxies, in order to eliminate the impact of unresolved velocity gradients. This makes the determination of a local intrinsic line width in clumps near the center and in compact galaxies often tricky and unreliable. A case in point is BX599, where the observed dispersion toward clump A definitely is broadened to an effective local σ of 125 km/s. For the reasons discussed just before, in Figure 10 we use $\sigma_0$ ~76±20 km/s, which is an estimate from the line width outside this bright clump.

With the possible exception of extreme clumps, local star formation feedback thus does not appear to directly drive the local rms-velocity dispersion of the ionized gas. While the star formation-driven galactic outflows discussed in the last section are energetically capable of stirring up the ionized gas in the disk, it appears that the ordered outflows do manage to break out of the local environment. This is consistent with the observations of Steidel et al. (2010) who find that the outflowing gas is transported to ≥100 kpc. For this breakout to be efficient, the clumps probably must have clumpy structure below our current resolution, with a small filling factor of the densest gas. Such substructure is hinted at from a comparison of the velocity channel maps of the bright clump A in BX482, shown in the middle two rows of Figure 1. There are significant small scale variations for clump A in these individual velocity maps, consistent with spatial-velocity substructure on sub-kpc scales.

The question remains whether the kinematics of the ionized gas is a good proxy of that of the entire cold (molecular) gas in high-z SFGs. This important issue can soon be addressed with high resolution millimeter interferometric imaging of CO rotation emission lines (see Tacconi et al. 2010, Daddi et al. 2010a).

### 3.4 Are the clumps rotationally supported?

Most of the available numerical simulations of the z>1 gas rich disks predict that the gravitationally unstable clumps contract, spin-up and may approach a Jeans equilibrium with half or more of the support in rotation (Immeli et al. 2004 a,b, Dekel et al. 2009b, Agertz et al. 2010, Aumer et al. 2010, Ceverino, Dekel & Bournaud 2010). This is because in these simulations the angular momenta of the collapsing clumps are largely conserved.

We have explored the evidence for rotation in our data by determining the velocity gradients across clumps in the 'raw' and 'residual' velocity maps. Figures 11 and 12 show the residual velocity distributions in BX482, ZC406690 and D3a15504, after subtracting (by kinemetry or modeling) the large scale velocity gradients caused by the overall galaxy rotation. Clump rotation should show up as a local gradient in these residual maps. If the clumps originally have a similar angular momentum direction as the galaxy, their rotation should be prograde. For this reason, the right panels in Figures 11 and 12 show position- velocity residual cuts through several of the largest and best isolated clumps, along the maximum velocity gradient ('line of nodes') of the galactic rotation, where the largest effects are expected in most cases. We have also explored other directions, with little difference in the results.

Velocity gradients are indeed present in the velocity maps across the clumps. In the 'raw' maps they are on average comparable in magnitude and sign ($\delta_{raw} = \frac{(v_{\max} - v_{\min})_{raw}}{2\sin(i)R_{clump}} \sim 30\ (\pm 11)\ \ \mathrm{km/s/kpc}$) to the large scale velocity gradients across the galaxies but there are no large *additional* local gradients. In D3a15504 caution is warranted as the largest gradients (through clumps C, E and F) may also be interpreted as large scale, radial streaming of the circum-nuclear gas in a barred potential, as discussed in Genzel et al. (2006). After subtraction of the large scale velocity gradients from galaxy rotation, the inclination corrected 'residual' velocity gradients typically are $\delta_{residual} = \frac{(v_{\max} - v_{\min})_{residual}}{2\sin(i)R_{clump}} \sim \pm 15\ (\pm 5)\ \ \mathrm{km/s/kpc}$. These residual gradients are often retro-grade (negative sign in row 31 of Table 2).

Are these observed gradients consistent with the clump mass estimates (row 8 of Table 2) in virial equilibrium? In principle, dynamical masses can be computed for rotationally supported systems if rotation velocity and inclination are known. However, most of the giant clumps have HWHM radii comparable to or slightly larger than our resolution, so that beam smearing plays an important role in lowering the expected velocity gradients.

To get a quantitative handle on how large these resolution effects are we took two approaches. In the first we constructed simple toy models of rotationally supported clumps of different masses ($5\mathrm{x}10^{8}$ to $10^{10}\ M_{\odot}$), sizes (HWHM radii from 0.2 to 1.7 kpc) and intrinsic density distributions (Gaussian or uniform). For these model clumps we calculated model data cubes from the input mass distributions, for a range of inclinations and masses, and for a z-thickness of $<z>/<R>\sim 0.2$ appropriate for the high-z galaxies. We then convolved these models with the spatial (~0.2" FWHM) and spectral (85 km/s FHWM) instrumental resolution, and sampled them at the pixel scales of SINFONI. The left panel of Figure 13 shows the ratio of the clump mass to the empirical 'rotational' dynamical mass of these model clumps, given by

$$M_{dyn\ rot} \sim b\beta\left(\frac{v_{\max}-v_{\min}}{2\sin(i)}\right)^2 \frac{R}{G} \sim b\ \beta\ 2.3\cdot 10^5\, \delta^2_{km/s/kpc} R^3_{kpc} \quad (M_\odot) \qquad (6).$$

Here δ is the observed velocity gradient across the observed size of a clump (2R), after correction for the inclination of the galaxy, and b and $\beta$ are dimensionless numbers; β is dependent on the assumed density distribution ($\beta$=1 and 1.16 for uniform and Gaussian clumps, for instance). Figure 13 shows that mass estimates with equation (6) require an average value of b=4.4 (with substantial scatter) for matching the toy model clump masses if $R=R_{HWHM\ obs}$ and if clump sizes are comparable to or larger than the beam sizes. This is mainly because the true 'Keplerian' limit is only reached at R~1.5-2 $R_{HWHM}$. This can be best seen from the open symbols in the left panel of Figure 13, which shows that for the same basic assumptions b approaches unity for R=2 $R_{HWHM}$. In the following we will use the estimator at $R=R_{HWHM}$ since in the real data the confusing effects of background and other clumps are less severe close to the clumps' cores.

In our second approach we analyzed the properties of several prominent clumps in a $M_{baryon}=3\times10^{10}\ M_\odot$ galaxy in the cosmological adaptive mesh, hydro-simulations of Ceverino et al. (2010, and in prep.), to which publications we refer the reader for more details. The simulations have a resolution of 35-70 pc. At z=2.3 the simulated galaxy has a number of clumps of mass ~$3\times10^8\ M_\odot$, radius ~0.3 kpc and intrinsic circular velocity of 70 km/s. These clumps are largely rotationally supported. From the simulation, we constructed data cubes of the Hα emission at different resolutions and inclinations, which we then analyzed in the same manner as for SINS data cubes. At the resolution of our SINS data, the clumps in the simulated galaxy exhibit clear rotational signatures, and we extracted velocity gradients in the same manner as for the real SINS data. Since the intrinsic clump masses are known, it is then possible to compute b in equation (6). The simulated galaxy has much smaller galaxy and clump masses and radii than the SINS galaxies. We expect the clump quantities to scale with the galaxy properties, and $v_{clump, rot} \sim R_{clump} \sim (M_{clump})^{1/3} \sim (M_{disk})^{1/3}$ (equation 5). So for clumps 10 times more massive (as in the more massive clumps in our sample), we expect the size and velocity of the typical clumps to be twice as large as the clumps in the simulations. In order to refer to clumps twice as extended as the simulated clumps, we can pretend that the beam smearing is half the true value, namely FWHM=0.1"=0.84 kpc. We read δ with that smoothing, and don't need to correct this value because V and R scale similarly. The filled magenta triangle and open black circle in Figure 13 give the average and scatter of four clumps in the simulated galaxy, scaled in this manner to clump masses of $3x10^9$ $M_\odot$ and $10^{10}$ $M_\odot$, respectively. The inferred calibrations for b from the simulated galaxy are in excellent agreement with those from the toy model clumps.

With this calibration (b=4.4) for rotationally supported clumps, we then proceeded to compute dynamical clump masses for the SINS clumps in Table 2, from the observed 'raw' velocity gradients. The inferred 'rotational' dynamical masses range between $2x10^8$ and $2x10^9$ $M_\odot$ (row 32 of Table 2). As before, the gas masses were estimated from equation 2. The SINS galaxies have clump gas masses on average 6.4 times larger (with a wide range from 3 to 75) than the rotational dynamical masses estimated from their velocity gradients with b=4.4. Including stellar masses further increases the discrepancy. For instance the stellar mass of clump A in BX482 A is ~3x10[9] $M_\odot$ (Förster Schreiber et al. 2011b, in prep.), so that the total clump mass is 1.4 times the gas mass. The discrepancy increases by another factor of ~2 if 'residual' velocity gradients are used for computing dynamical masses.

How can one understand this discrepancy? Extended Hα emission surrounding the clump may decrease the rotational signal. If the clumps were much smaller than our resolution the rotational signal would be completely washed out by beam smearing (Figure 13). In that case equation (6) with b=4.4 would underestimate the dynamical masses significantly. However, the clumps should then in turn exhibit very large central velocity dispersions (100-170 km/s) caused by the same beam smearing.

Such large velocity dispersion maxima centered on the clumps are not observed in our sample, although the data shown in Figures 10, 15 and 16 are consistent with a modest increase in velocity dispersion toward the clump centers. Genel et al. (2010) and Aumer et al. (2010) find in their simulations that clumps actually are minima in the galaxy wide velocity dispersions, and argue that turbulence created on large scales is dissipated more efficiently in the dense clump environment. If this were to apply to our SFGs, a significant amount of beam-smeared rotation could 'hide' in the current velocity dispersion maps.

If the clumps are supported by a combination of rotation and pressure the appropriate measure for the dynamical mass is

$$M_{dyn\ r+p} = 2.33x10^5 R_{kpc} \beta d \left(2\sigma^2_{km/s} + \delta^2_{km/s/kpc} R^2_{kpc}\right) \qquad (M_\odot) \qquad (7),$$

where $\beta$ is defined as in equation (6), and d is a dimensionless calibration factor. Our clump modeling suggest d=0.4-1.2. For d=1 and a clump velocity dispersion $\sigma_0$ between 50 and 95 km/s, the pressure term ranges between 1.5 and $5 \times 10^9$ $M_\odot$ for the different clumps in Table 2, and appears to dominate over the rotational mass. Equation (7) then yields an average ratio of gas mass to dynamical mass of ~2d, and somewhat larger if stellar masses are included (row 35 of Table 2). For d=1 gas (plus stellar) masses still exceed the dynamical masses but for d=0.5 the estimates agree to within the uncertainties. The application of equation (7) to the SINS clumps may thus give a reasonable match to the estimated gas and stellar masses in these clumps, always keeping in mind the very substantial uncertainties in all these estimates. Note that the estimator (7) also works well for rotationally supported but unresolved clumps.

The inferred discrepancy between the rotational masses from equation (6) and the gas masses is most strongly driven by the prominent clumps A in BX482 and ZC406690 (Figure 11). Even when allowing for large uncertainties, the relatively small observed velocity gradients in these two cases appear inconsistent with rotational support. These clumps may either be largely pressure supported, with rotation contributing perhaps 10-20% of the energy, or they are not virialized (Genel et al. 2010). In the smaller and probably more typical clumps of D3a15504 and ZC782941 rotational support is more significant and may even dominate, considering the uncertainties in our analysis, limitations in signal to noise ratio and the possible contamination by unrelated background emission. This is especially true if most of the mass and rotation of the clumps are on scales much smaller than our beam, and at the same time the clumps are local minima in the galaxy wide velocity dispersion.

There could also be physical reasons for low rotational velocities. The first is that the clouds may not be undergoing global collapse. Angular momentum may not be conserved due to outward transport by large scale torques or magnetic fields. Milky Way GMCs also typically have little rotation (δ~45 km/s/kpc: Blitz 1993, Phillips 1992), similar to the values we find in our SINS clumps. A fraction of these GMCs also have a retrograde velocity gradient with respect to the rotation of the Milky Way (Blitz 1993), inconsistent with the simple spin-up scenario from initially differentially rotating disk gas. The most extreme clumps may also not have enough time for virialization before much of the gas is expelled, as is suggested by the simulations of Genel et al. (2010).

### 3.5 Is the velocity dispersion isotropic?

If the galactic turbulence is created by clump-clump interactions, the velocity dispersion may be anisotropic, with larger dispersion in the galactic plane than perpendicular to it, as in the simulations of Aumer et al. (2010). Figure 14 is an attempt to test this prediction for those well resolved disk galaxies within the SINS sample (Förster Schreiber et al. 2009) and other recent integral field data (Wright et al. 2007, van Starkenburg et al. 2008, Epinat et al. 2009, Lemoine-Busserolle & Lamareille 2010), where a reasonably robust value of the inclination is known from dynamical fitting (Cresci et al. 2009), or from the geometric aspect ratio of the emission.

Keeping in mind the large uncertainties in $sin(i)$ ($\Delta\, sin(i) \sim \pm 0.15$, including systematic effects), there is a tantalizing trend for the more edge-on systems to have relatively larger line-of-sight dispersion $\sigma_0$, or smaller $v_c/\sigma_0$ (see also Aumer et al. 2010). Dividing up the data into two bins ($sin(i)$>0.64 and <0.64), the two sets differ at ~2-2.7σ. If real this trend would correspond to an anisotropy in the velocity dispersion ellipsoid of $\sigma_{plane}/\sigma_{pole}$~2 (grey curve in Figure 14).

Two cautionary remarks are in order. One is that only two of the highest inclination disks (*sin(i)*>0.8) have AO data sets, and these two have fairly low velocity dispersion. The highest inclination (and high quality) disk BX389 (Cresci et al. 2009, Genzel et al. 2008) does have a very high velocity dispersion ($\sigma_0$=87±10 km/s) but this result comes from seeing limited data. The other is that in a highly inclined disk velocity dispersion and rotation are more entangled than in a more face-on system. As a result the observed velocity dispersion in modest resolution data may be artificially increased by unresolved rotation and confusion of different regions along the line of sight. Given these concerns, we consider the trend in Figure 23 tantalizing but not convincing.

### 3.6 Are there spatial variations in chemical abundances ?

The flux ratio 6585 Å [NII])/Hα in non-AGN SFGs is a measure of the oxygen to hydrogen abundance ratio (Pettini & Pagel 2004, Erb et al. 2006a). It is thus interesting to ask whether the intense star forming activity in the giant clumps is reflected in local enhancements in metallicity, and whether there are abundance differences between the central and outer parts of the galaxies because of the expected inside-out evolution of the galaxies (e.g. Somerville et al. 2008).

We first analyzed specific regions from extracted spectra in order to maximize the SNR, with the quantitative results listed in Table 3. Very significant variations in [NII]/Hα are indeed present in ZC406690. This is easily seen by inspection of the spectra in Figure 7. The [NII] line is weak toward clump A or region D, while it is much stronger toward clumps B and C. In terms of inferred oxygen abundance (Table 3), these changes correspond to an increase of more than a factor of 2 from $Z_O$~0.4 $Z_\odot$ in clump A to $Z_O$~0.70 $Z_\odot$ in clump B. Clumps A, B and C also form a downward sequence in the ratio of Hα line flux to rest-frame 2700Å continuum flux density from the I-band imaging, which may be explained by a combination of increasing extinction and increasing age. The observations in ZC406690 may thus suggest a trend of increasing heavy element abundance with age of the star forming clumps in the young star forming ring of this galaxy.

In the comparably young star forming ring in BX482 (Genzel et al. 2008, Förster Schreiber et al. 2011b, in preparation) clumps A to C and the rest of the Hα-ring have similar abundances while the central 'cavity' near the kinematic center has an [NII]/Hα ratio 2.7σ larger. The inferred oxygen abundance increases from $Z_O$~0.56 $Z_\odot$ in the bright Hα star forming ring to $Z_O$~0.73 $Z_\odot$ near the center. Correspondingly Förster Schreiber et al. (2011a, and 2011b, in preparation, section 4.2.2) find an age gradient from <100 Myrs in the ring clumps, to >200 Myrs near the center. In the somewhat more mature galaxy ZC782941 the clumps near the kinematic center of the galaxy (B-E) have an [NII]/Hα ratio 3.3σ times greater than in the outer clump A (or minor merger) and the interclump regions, corresponding to an increase from 0.67 to 0.84 $Z_\odot$.

Finally, the most mature galaxy (in terms of stellar age and abundances) in our sample, BzK15504, does not show any evidence for a difference between clumps and interclump gas in the main body of the disk. In our AO data the [NII] line is too faint in the outer parts of the galaxy disk for reliable statements on the oxygen abundance. However, on the basis of seeing limited data of D3a15504, which are sensitive to lower surface brightness, Buschkamp et al. (in prep.) deduce ~0.3-0.5 $Z_\odot$ abundances in the outer disk. The circum-nuclear region has a higher [NII]/Hα, corresponding to an increase in oxygen abundance from 0.9 $Z_\odot$ in the disk to 1.07 $Z_\odot$ near the nucleus.

In agreement with our findings, Buschkamp et al. (in prep) find that [NII]/Hα increases from the outer regions toward the nuclei in two additional massive and mature SFGs, BX610 (z=2.2, see also Förster Schreiber et al. 2006) and BzK6004 (z=2.4). In these cases the inferred abundances appear to increase from slightly below solar (~0.95 $Z_\odot$) in the outer disk to super-solar (~1.2 $Z_\odot$) near the center. In all these cases one needs to caution that such line ratio gradients may be caused by the presence of a central AGN. Indeed this explanation may apply to BzK15504 (Genzel et al. 2006). Evidence for a weak AGN causing a radial change in [NII]/Hα and [OIII]/Hβ line ratios has also been found for another z~1.5 SFG by Wright et al. (2010).

We have also explored the data of BX482, BzK15504 and ZC782941 in more detail by analyzing the [NII]/Hα pixel to pixel variations statistically and by searching for correlations between [NII]/Hα and $\Sigma_{\rm star\ form}$. This analysis confirms the results on selected regions that we discussed in the last paragraph, but does not reveal significant additional trends. This is perhaps not surprising, since even with the superior data presented in this paper, the typical average SNR per pixel in the [NII] line is modest.

In summary of this section, we find evidence for clump to clump and center to outer disk variations in the inferred gas phase oxygen abundance that are qualitatively consistent with the expected rapid heavy element enrichment in the young disks. The radial abundance gradients are broadly consistent with inside-out growth predicted from semi-analytic models (e.g. Somerville et al. 2008). These inside-out gradients have the opposite trend of the gradients inferred by Cresci et al. (2010) for three z~3 Lyman break galaxies, which appear to have minima of metallicity towards the brightest [OIII] emission line peaks near the galaxy centers, perhaps as the result of recent accretion of fresh, low metallicity gas.

# 4. Discussion

The most important results of the last section can be summarized as follows,

- The giant star forming clumps in the observed z~2 SFGs reside in regions where the Toomre *Q*-parameter is at or below unity. The giant clumps are thus plausibly bound and the high-z disks we have studied are unstable against gravitational collapse on galactic scales;
- the giant clumps are the launching sites of powerful outflows, probably driven by the energy and momentum released by massive stars and supernovae in the clumps;
- velocity gradients across clumps are modest (10-40 km/s/kpc) and comparable to the large scale galactic gradients. Given beam smearing effects, finite signal to noise ratios and contamination by diffuse emission, typical clumps may still be rotationally supported, but extreme clumps may not be. These may either be predominantly pressure supported, or they are not virialized;
- the large velocity dispersions are spatially fairly constant. With the exception of the extreme clumps in ZC406690, the clumps leave little or only relatively small local imprints on this distributed 'floor' of galactic scale, local random motions ('galactic turbulence'). The dependence of the amplitude of this floor on galactic or local star formation surface density is modest. The change in velocity dispersion across more than an order of magnitude in star formation surface density is less than a factor of two and comparable to the scatter of the

- there are modest but significant clump to clump, and center to outer disk variations of the [NII]/Hα ratios. Keeping in mind the possible contaminations from central AGNs, these variations are broadly consistent with active nucleosynthesis in inside-out growing disks, and/or with clump migration in a disk that is fed semi-continuously with fresh gas from the halo.

In the following section we analyze the impact of these findings on the issues of star formation feedback and galactic 'turbulence'. We begin with a quantitative analysis of the derived mass outflow rates, followed by discussions of the implied lifetimes and evolution of the giant clumps and the origin of their turbulence.

## *4.1 Estimates of mass outflow rates*

### 4.1.1 modeling the Hα emission in the context of current models of galactic winds

Previous work on z~0 galactic winds is based to a significant extent on observations of edge-on 'starburst' galaxies with optical emission lines, soft and hard X-ray emission, blueshifted interstellar absorption lines, outflowing molecular gas or a nonthermal radio continuum halo (Heckman, Armus and Miley 1990, Lehnert & Heckman 1996, Walter, Weiss & Scoville 2002, Heckman 2003, Veilleux et al. 2005, Strickland & Heckman 2009). According to the 'standard' model, the winds are set up by supernovae and stellar winds injecting kinetic energy and momentum into the surrounding medium and creating an over-pressured bubble of hot gas. This bubble expands and sweeps up the ambient medium (Heckman 2003, Veilleux et al. 2005). An alternative or additional mechanism is the effect of radiation pressure onto dust grains, creating a 'tepid' but massive, cool outflow in luminous, 'super-Eddington' galaxies (Murray et al. 2005, Martin 2005, Murray et al. 2010a). This model may explain the empirical linear scaling of outflow velocities with galaxy circular velocities in z~0 ultra-luminous infrared galaxies (ULIRGs, $\Delta v_{max} / v_c$ ~1.5-3, Martin 2005, Veilleux et al. 2005).

In a stratified disk, the bubble will expand preferentially along the minor axis of the disk, and will eventually 'blow out' into the halo, although the magnitude and appearance of this blow out may be strongly modified and lowered in a strongly inhomogeneous interstellar medium (Cooper et al. 2008). The escaping wind fluid sweeps up, entrains and shocks ambient cloud material from the disk and halo (Veilleux et al. 2005 and references therein). Optical emission lines may originate in this shocked ambient medium. In nearby cases such as M82 Hα/[NII] emission tends to come from compact filaments and knots at the edge of the X-ray emission nebulae that may originate from internal shocks in the hot wind fluid and the interaction of the hot gas with the ambient medium (Heckman, Lehnert & Armus 1993, Westmoquette et al. 2007). Alternatively, the warm and cool dusty gas may initially be lifted to hundreds of parsecs by radiation pressure, and only then further accelerated by the hot wind component (Murray et al. 2010a).

If this standard model also applies to the outflows in z~2 SFGs, the broad Hα/[NII] emission is well suited for quantifying the mass, outflow rate, momentum and energy in the warm/cool component of the high-z winds. The fundamental advantages of the Hα measurements presented above for estimating mass outflow rates are that they directly yield outflow velocities, sizes and – if local electron densities are known – also hydrogen column densities/masses, without the need for an assumption of heavy element abundances, or correction for line radiative transport. The greatest disadvantage of Hα is that the line luminosity is proportional to 'emission measure' ($\propto \int n_e^2 dV$). As a result the derived mass outflow rates are sensitive to the adopted radial density structure but not to the geometry (e.g. opening angle) of the outflow. The Hα emission is obviously also sensitive to dust extinction in front of and mixed in with the outflow. Such 'wind' dust is clearly seen in the M82 outflow, for instance (Roussel et al. 2010). In our analysis we have corrected for foreground extinction derived from the UV/optical SED fitting of our galaxies (row 5 in Table 2). The preferentially blue-shifted broad wings show that differential extinction along the line of sight does indeed play an important role for the interpretation (section 3.2.1). Our current observations do not give any handle on how this internal dust is distributed. We thus adopt first order correction factors ranging between 1 and 2 (row 16 of Table 2) depending on the asymmetry and blue-shift of the broad line emission.

In analogy to z~0 winds, the most likely excitation mechanism for the Hα emission is photoionization from the central ionization sources, or from hot, postshock gas in wind-gas collisions. If the gas is photoionized the data can be analyzed with standard case B recombination theory (Osterbock 1989). In Appendix B we present our detailed analysis with two different modeling assumptions on the density structure of the flow. In principle it may also be possible that the wind or postshock gas is hot enough that collisional ionization plays a role, although the evidence in z~0 starburst galaxies suggests that this mechanism is unlikely to dominate the overall excitation. For completeness we also derive in Appendix B the mass loss rates for pure collisional ionization, which turn out to be a factor ≤ 2 smaller than the recombination values. Table 2 (rows 20 to 23) lists the derived outflow rates.

### 4.1.2 ionized outflow rates in clumps exceed star formation rates

Our main finding is that the outflow rates in ionized gas are as large as or larger than the star formation rates. In the most extreme clump B in ZC406690 we infer an outflow rate 8.4 times greater than the star formation rate. Depending on the extinction corrections, geometries, modeling assumptions etc., the exact values of the derived outflow rates are probably uncertain by a factor of at least 3 (up, and down). Our findings are in very good agreement with the conclusions of Erb et al. (2006a) and Erb (2008) who deduce ratios of outflow to star formation rates between 1 and 4 from an analysis of the z~2 stellar mass-metallicity and stellar mass-star formation relations.

### 4.1.3 correction for other gas phases

These outflow estimates are probably lower limits, since in addition to the warm ionized gas, there very likely are substantial components of very hot, as well as cold (atomic and molecular) gas. The latter may be especially important if radiation pressure onto dust grains in the outflows plays an important role, as the dusty molecular gas is coupled strongly to UV radiation. Unfortunately the relative contributions of the different gas phases are not well known even in z~0 SFGs. In the relatively low-mass starburst galaxy M82 the mass of atomic hydrogen in the outflowing nebula is estimated to be about $5x10^7$ $M_{\odot}$, which is comparable to the mass of hotter gas producing X-rays (Heckman, Armus and Miley 1990, Crutcher,

Rogstad and Chu 1978, Stark and Carlson 1984, Nakai et al. 1987, Watson, Stanger and Griffiths 1984, Schaaf et al. 1989, Fabbiano 1988), and about six times smaller than the amount of molecular gas associated with the outflow (~$3x10^8$ $M_\odot$, Walter et al. 2002) . In contrast the mass of material responsible for optical line emission may be much smaller, ~ $2x10^5$ $M_\odot$. The warm and cold components in M82 move at roughly the same velocity, whereas the hot wind fluid has a velocity around ten times larger. Thus if one would scale directly from these observations of M82, the winds in z~2 SFGs would have mass outflow rates 2 to 3 orders of magnitudes greater than what is measured with Hα alone. It would seem highly implausible that this could apply to our z~2 SFGs.

However, the structure of galactic winds in relatively low mass starburst galaxies (such as M82) at z~0 could very well differ greatly from that of more luminous, denser and more massive galaxies at both low-z (such as ULIRGs) and high-z, altering the relative contributions of the different wind components and drivers (supernovae vs. stellar winds and radiation). Weiner et al. (2009) has shown that powerful galactic winds are a common feature of most or all massive, luminous SFGs at z~1, while at z~0 they are only found in the relatively rare starburst and actively merging galaxies. One possible explanation for these differences may be that supernova driven hot gas may dominate in low mass galaxies, while cooler, radiation pressure driven winds may dominate in massive galaxies. In rows 36 & 37 of Table 2 we compare the force of the outflows to that delivered by radiation. The ratios range from a few to about twenty (in clump B of ZC406690). Values at the lower end of this range could thus plausibly be understood in the radiation driven wind picture, while values in the upper range would require very large optical depths and multiple scattering of infrared photons (Murray et al.2010a,b).

### 4.2 Clump lifetimes and evolution

#### 4.2.1 gas expulsion time scale

What do our observations tell us about the lifetimes of the giant clumps? We address this question in several ways. A first empirical estimate of the effect of gas expulsion by the outflows comes from the ratio of the molecular gas masses of the clumps to the outflow rates, with an upward correction by a factor of two based on the assumption that on average a clump is seen halfway through its evolution. There may also be atomic gas associated with the clumps but given their large gas columns (>>$10^2$ $M_{\odot}pc^{-2}$) that atomic fraction is probably small (Blitz & Rosolowsky 2006). The gas expulsion times estimated in this manner (row 24 in Table 2) range between 170 and 1600 Myrs.

#### 4.2.2 stellar ages

A second estimate for clump lifetimes can be made on the basis of the stellar ages of clumps estimated from population synthesis modeling. Elmegreen et al. (2009, see also Elmegreen & Elmegreen 2005) fitted constant star formation and exponentially decaying models to 4-6-band ACS/NICMOS photometry of clump-cluster and chain galaxies in the Hubble Ultra Deep Field. From this analysis they infer clump ages in z~2-3 clump-cluster and chain galaxies ranging widely between 10 and a few $10^2$ Myrs, with typical clumps having stellar ages of 100 to 200 Myrs and an upper envelope of about 300 Myrs. In the case of BX482 and D3a15504 it is possible to use the Hα-equivalent width as an age indicator that is fairly insensitive to extinction (Förster Schreiber et al. 2011b, in prep.). The analysis requires the subtraction of the interclump background. For the bright clumps A-C in BX482 the result is fairly immune to the treatment of the background and yields fairly robust ages between 30 and 100 Myrs (Förster Schreiber et al. 2011b, in prep.). In the case of D3a15504 the clumps are not apparent in the fairly low SNR NACO $K_s$-image and even in the Hα-image the background is a significant contribution to the clump aperture fluxes. Only estimates based on integrated photometry at the location of clumps (i.e. without subtraction of the underlying host galaxy's light) can be reliably obtained. Inferred ages are in excess of ~ 1 Gyr, and represent upper limits given the contribution of the host galaxy's stellar population. For ZC782941 and ZC406690 the ACS I-band images sample the rest-frame UV, and the ratio of Hα to I-band flux density (restframe 2700 Å) is very sensitive to extinction. No reliable ages can be deduced but we have estimated age ranges (80 to 800 Myrs for clumps A and B in ZC406690) based on various plausible star formation histories and extinctions. The deduced clump stellar ages are listed in row 25 of Table 2.

The best available estimates of stellar ages of the giant clumps, while very uncertain, thus may set an upper limit to the clumps' stellar ages, and thus plausibly their lifetimes, of ~300 Myrs. A cautionary remark is that the stellar lifetimes so deduced really sample only the average ages of the stars dominating the stellar (UV plus optical) light, and thus may underestimate the true ages of the star forming clumps if there is an underlying older population (Maraston et al. 2010). In addition the assumption of constant star formation histories may introduce systematic uncertainties in z~2 galaxies with ongoing large gas accretion (Maraston et al. 2010).

### 4.2.3 metal enrichement time scale

Thirdly, we can also use the [NII]/Hα-ratio and the inferred oxygen abundances for obtaining upper limits to the clump ages. Given the [NII]/Hα values in Table 3 we calculate the metallicity using the prescription of Pettini and Pagel (2004) and, assuming the solar ratio of mass in oxygen to total mass in heavy elements, derive the total mass fraction of heavy elements (Z). Assuming an initially zero metallicity, we use both a "closed-box" and "leaky box" model (Erb 2008) to determine the time required to produce the observed Z. The derivation of the timescale is discussed in more detail in Appendix C. The closed box estimates range between 150 Myrs and 900 Myrs (row 26 in Table 2).

The timescales derived with the "leaky box" model are all longer than those of the "closed box" model, ranging from comparable ages to ones about three times longer (see row 27 in Table 2), with one outlier with an unrealistic 'leaky box' enrichment time scale. There are a few objects with very long metal enrichment timescales in the leaky box model. The aforementioned outlier could arise from measured SFRs that are much lower than in the past. As these estimates assume that the gas forming the clumps begins with zero metal mass fraction and therefore that all of the observed metals currently in the clumps were produced in them, they tell us about the maximum possible age of the clumps. The real metal enrichment time scales are smaller, since much of the gas in the clumps was likely somewhat enriched when the clumps formed. To explore this idea further, we have calculated the time required to enrich clump A of ZC406690 to the metallicity of clump B in the same galaxy. Clump A is the brightest, highest Hα equivalent width, and thus likely the youngest clump in ZC406690, and therefore could be representative of the initial metallicity of forming clumps. On the other hand, clump B appears to be much more evolved and potentially past its peak in SFR. With the closed model, this timescale is 130 Myr, and 210 Myr with the leaky model. These are more convincing ages for evolved clumps like clump B, since they don't assume the clumps formed from zero metallicity gas.

### 4.2.4 expansion time scale

Finally, we consider the Taylor-Sedov solution for an explosion in the radiative phase to estimate the time scale for the outflow to expand to its current size. In the Taylor-Sedov solution (Osterbrock 1989), the radius of a shock is given by

$$R_s = 12.8\left(\frac{t_{\exp}}{10^4}\right)^{2/5}\left(\frac{E_{51}}{n_0}\right)^{1/5} pc \qquad (8),$$

where $t_{exp}$ is the time since the explosion in years, $E_{51}$ is the energy released by the SN in units of $10^{51}$ ergs, and $n_0 \sim 10^{-2} cm^{-3}$ is the ambient ISM density (McKee and Ostriker 1977). We take $R_s$ as the radius of the clump, assuming the expanding superbubble still follows this relation in the plane of the disk, even though the bubble has blown out along the minor axis. We estimate the input energy using a simple order of magnitude calculation,

$$E = \dot{M} v_{out}^2 t_{\exp} \varepsilon_{ff} \qquad (9),$$

where we use the outflow velocities and rates listed in Table 2 and $\varepsilon_{ff} \sim 0.1$ is an efficiency factor which takes into account the fact that most of the energy is released perpendicular to the disk and not in the plane of the disk.. The resulting timescales are listed in row 28 of Table 2 and constitute lower limits to the clump life times. The values for $t_{exp}$ in Table 2 have a median of 130 Myrs.

### 4.2.5 average lifetime and fate of the clumps

The discussion in the last sections suggests that the typical lifetimes of the giant gaseous clumps are probably less than 1 Gyr, and on average perhaps closer to 500 Myrs, with large uncertainties and scatter. The powerful outflows probably play an important factor in setting these lifetimes. While the measurements are probably still too uncertain to be sure of the final outcome(s), less active clumps (BX482 A or the clumps in D3a15504) may convert approximately half of their initial gas to stars (row 30 in Table 2) and remain basically intact. Given that the in-spiral time by dynamical friction is also less than 500 Myrs (equation 1, Noguchi 1999, Immeli et al. 2004a,b, Dekel et al. 2009b, Ceverino et al. 2010), these clumps could plausibly migrate all the way to the nuclear region (row 29 in Table 2). In the most active and compact clumps (such as ZC406690 A and B), however, the feedback may disrupt the clumps before they have converted a substantial fraction of their initial gas to stars. Such clumps may disperse in the outer disk before having had time to migrate into the center. The dissolving stellar component of the clump may still continue the in-spiral and contribute to the secular buildup of a central bulge, albeit with smaller efficiency than assumed in the original work of Noguchi (1999) and Immeli et al. (2004 a,b). An interesting question is whether globular clusters might be formed in this environment (Shapiro, Genzel & Förster Schreiber 2010) and if so, how they would survive the strong feedback.

Our findings are in broad agreement with an ‘average’ of the current analytical estimates (Murray et al. 2010a,b, Krumholz & Dekel 2010). The clump evolution in current hydro-simulations strongly depends on assumptions on sub-grid physics, reflecting the different discussion in the analytical work. None of these cosmological simulations come close yet to capturing the spatial structure and physical complexity of the interstellar gas physics (c.f. Wada & Norman 2001, Cooper et al. 2008). Exploring clump stability with different supernova energy feedback strengths Elmegreen et al. (2008) find that clumps live for a few hundred Myrs with the exception of a case with very strong feedback efficiency. Long clump lifetimes were also obtained by Ceverino et al. (2010) who also employed a supernova energy feedback scheme. In a recent simulation with radiation pressure, momentum feedback based on the Oppenheimer & Dave (2006) sub-grid recipes, Genel et al. (2010) find that the clumps are disrupted on a time scale of 50-100 Myrs, much more rapidly than indicated by the other theoretical estimates or our measurements. This is probably the result of their more extreme assumptions on the feedback efficiency, as compared to the original recipes of Oppenheimer and Dave (2006). As a result, the clumps in the Genel et al. (2010) simulation do not spiral inward much and do not have time to virialize. At the opposite extreme, even without winds, realistic clumps tend to be stripped by tidal forces and they loose maybe 50% of their mass before reaching the central bulge in 400-500 Myr (Bournaud et al. 2007).

### 4.3 Origin of the large turbulent velocities

What can our observations tell us about the origin of the large turbulent velocities that are characteristic of all $z\geq1$ SFGs?

The first obvious conclusion is that the velocity dispersion in the z~2 SFGs of our sample is not obviously driven by local stellar feedback. The near ‘constant’ floor of velocity dispersion may instead point to a galaxy-wide driving mechanism, such as the release of gravitational energy. Such an energy release could come from the inward clump migration in the disk and be driven by large scale torques or clump-clump collisions. Theoretical support for this mechanism is found in the simulations of Immeli et al. (2004 a,b), Agertz et al. (2009), Ceverino et al.(2010), and others. Alternatively gravitational energy might be released at the outer disk boundary by the accreting gas streams/flows from the halo. Support for this explanation comes from the simulations of Genel et al. (2010, and in prep.) who find that the most important driver of the local random motions is the accretion energy. Alternatively, and perhaps similar to the situation in the Milky Way, multiple agents may be at work simultaneously. The small spatial variations may then be the consequence of the fast communication speed of the drivers, which may include the fast outflows that couple to the disk over a wide range of scales (Elmegreen & Scalo 2004).

Recent detailed studies of the turbulent structure function in Milky Way GMCs also find little dependence of the cloud random motions on size/mass, environment and star formation activity of these clouds (Heyer & Brunt 2004, Brunt, Heyer & McLow 2009). Brunt et al. (2009) conclude that GMC turbulence is driven by a universal driver acting mostly on large scales, such as supernova-driven turbulence, magneto-rotational instability, or spiral shock forcing. According to these authors, small-scale driving by sources internal to molecular clouds, such as outflows, can be important on small scales, but cannot replicate the observed large-scale velocity fluctuations in the molecular interstellar medium.

A feedback dominated scenario may still be viable if the feedback ‘signal’ is hidden by other dependencies. In a self-gravitating disk with feedback-dominated turbulence the upward pressure generated by feedback is exactly compensated by the weight of the gas column above it. Ostriker & Shetty (2011) argue that in this case and for a Kennicutt (1998) gas-star formation relation the average gas density thus scales with star formation surface density and the z-velocity dispersion is approximately constant.

# 5. Conclusions

We have presented high quality SINFONI/VLT integral field spectroscopy of five massive, active star forming galaxies at z~2. Aided by natural and laser guide star adaptive optics, our data begin to constrain the nature and properties of the largest (diameter 1.5-3 kpc) and most massive (1-10x$10^9$ $M_\odot$), giant star forming clumps that are characteristic of many ‘normal’ high-z star forming galaxies.

We find clear evidence that the disks and giant star forming clumps are unstable to gravitational collapse and fragmentation. The giant star forming clumps plausibly represent the largest scale of gravitational instability in the high-z disks.

Spatially resolved, broad wings in Hα/[NII] emission associated with individual clumps make a strong case that the brightest giant star forming clumps launch massive galactic outflows driven by their intense star forming activity. We estimate that the mass outflow rates are comparable to or larger than the star formation rates, in one case by a factor of eight. These findings are in excellent agreement with other recent results on galactic outflows in high-z galaxies. The lifetimes of the giant clumps are strongly influenced and probably limited by these feedback effects to a few hundred million years but it is premature to be sure of the average clump evolution. For the less active clumps, a significant fraction of the original gas may be converted to stars at the end of this phase. Such clumps may even form bound entities that have a chance to migrate all the way to the nucleus and build up a bulge by secular effects. The most actively star forming clumps may, however, be destroyed before they have a chance to spiral much inward. In these cases stellar feedback and winds may thus limit the efficacy of the ‘violent disk instability’ mechanism for building central bulges. Our data suggest that the presence of strong outflows and star formation surface density are correlated, consistent with radiation pressure being important in driving the outflows.

The local rms-velocity dispersion of the ionized gas increases by less than a factor of 2 across a range of 20 in star formation surface density. This change is comparable to the scatter of velocity dispersions at any fixed star formation surface density throughout this range. To first order the large velocity dispersions in high-z galaxies are distributed as a spatially constant 'floor', similar to the situation in Milky Way GMCs. We also see only relatively modest intra-clump velocity gradients indicative of clump rotation. Our analysis suggests that two of the brightest giant clumps cannot be rotationally supported; these clumps may instead either be mainly pressure supported, or they are not virialized. Given our limited angular resolution and signal to noise ratios, the situation for most other clumps is less clear and in these cases rotational support may be significant or even dominant.

The absence of a strong correlation between velocity dispersion and star formation surface density suggests that strong local feedback is not the dominant driver of the large turbulent motions, or that the feedback drives this turbulence mainly on super-clump scales, unless secondary dependencies 'hide' the feedback signal. The diffuse and spatially constant nature of the random motions we observe may favor that the large scale release of gravitational energy is the key driver of the galactic 'turbulence'. This release could either occur within the disk as a result of the clumps' inward migration, or at the outer boundary of the disk as a result of the interaction with the gas streams and clumps/minor mergers coming in from the halo.

We find moderately significant evidence for clump-clump and center-outer disk variations in the inferred gas phase oxygen abundance. These variations are qualitatively consistent with the expected rapid heavy element enrichment in the young disks. The radial abundance gradients are broadly consistent with the expected inside-out growth of the young galaxies and the scenario of inward clump migration.

*Acknowledgements: We thank the staff of Paranal Observatory for their support. We also acknowledge the support of our other colleagues on the ESO Large Program 183.A-078: F.Bournaud, K.Caputi, A.Cimatti, E.Daddi, G. de Lucia, O.Gerhard, P.Johansson, S.Khochfar, O.LeFevre, V.Mainieri, H.McCracken, M.Mignoli and M.Scodeggio, as well as A.E.Shapley. We are grateful to the referee, Frederic Bournaud, for his comments and suggestions for improvement of our manuscript.*

# Figures

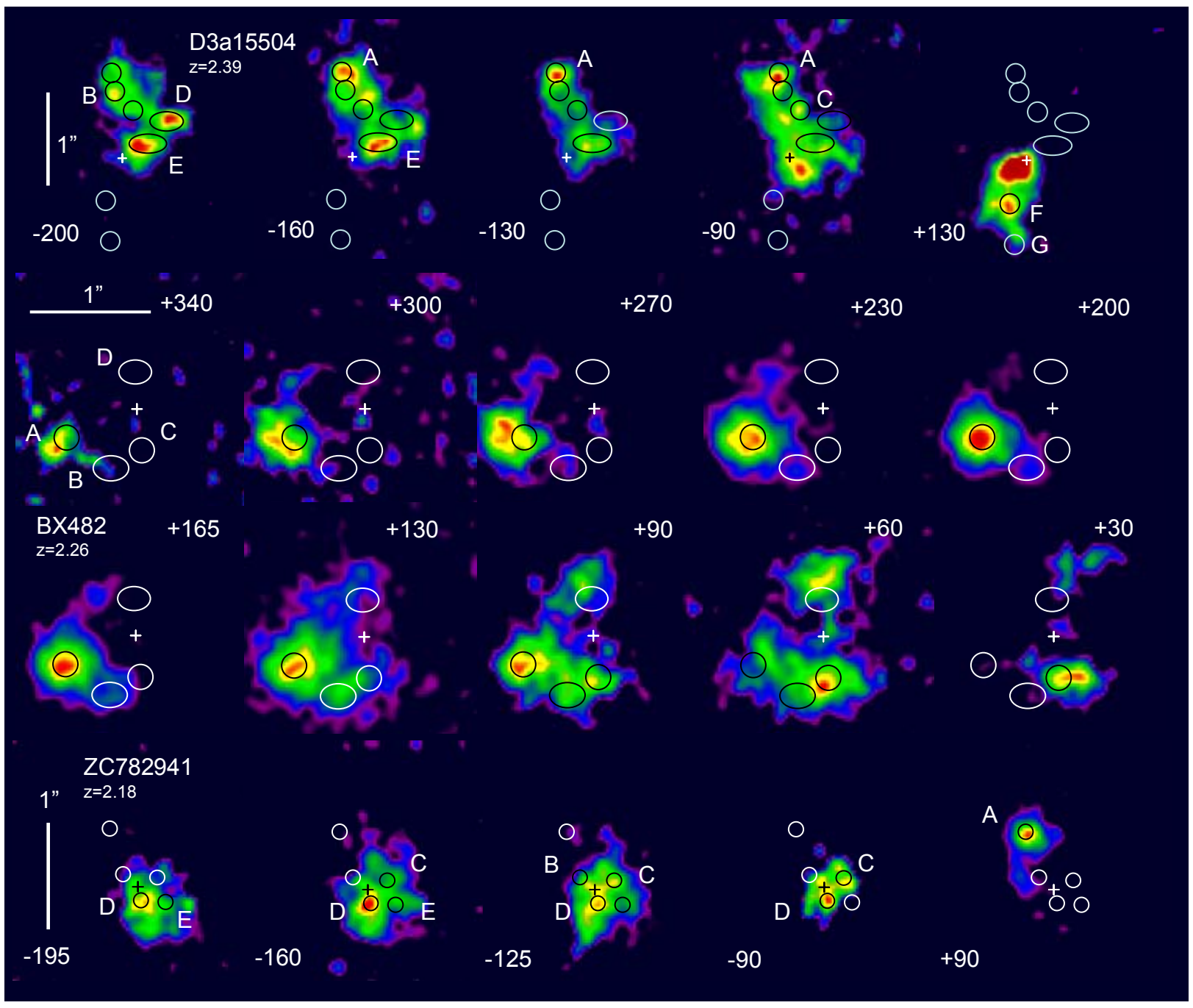

Figure 1. Maps of individual velocity ‘channels’ of width ~34 km/s in the Hα line of D3a15504 (top row), BX482 (middle two rows) and ZC782941 (bottom row). The maps are resampled to 0.025” per pixel and have a resolution of FWHM ~0.17-0.22”. Velocities relative to the systemic redshift indicated are given in km/s. Circles/ovals and symbols denote the clumps identified in these galaxies. Crosses denote the kinematic centers of the galaxy rotation. The color scale is linear and auto-scaled to the brightest emission in each channel.

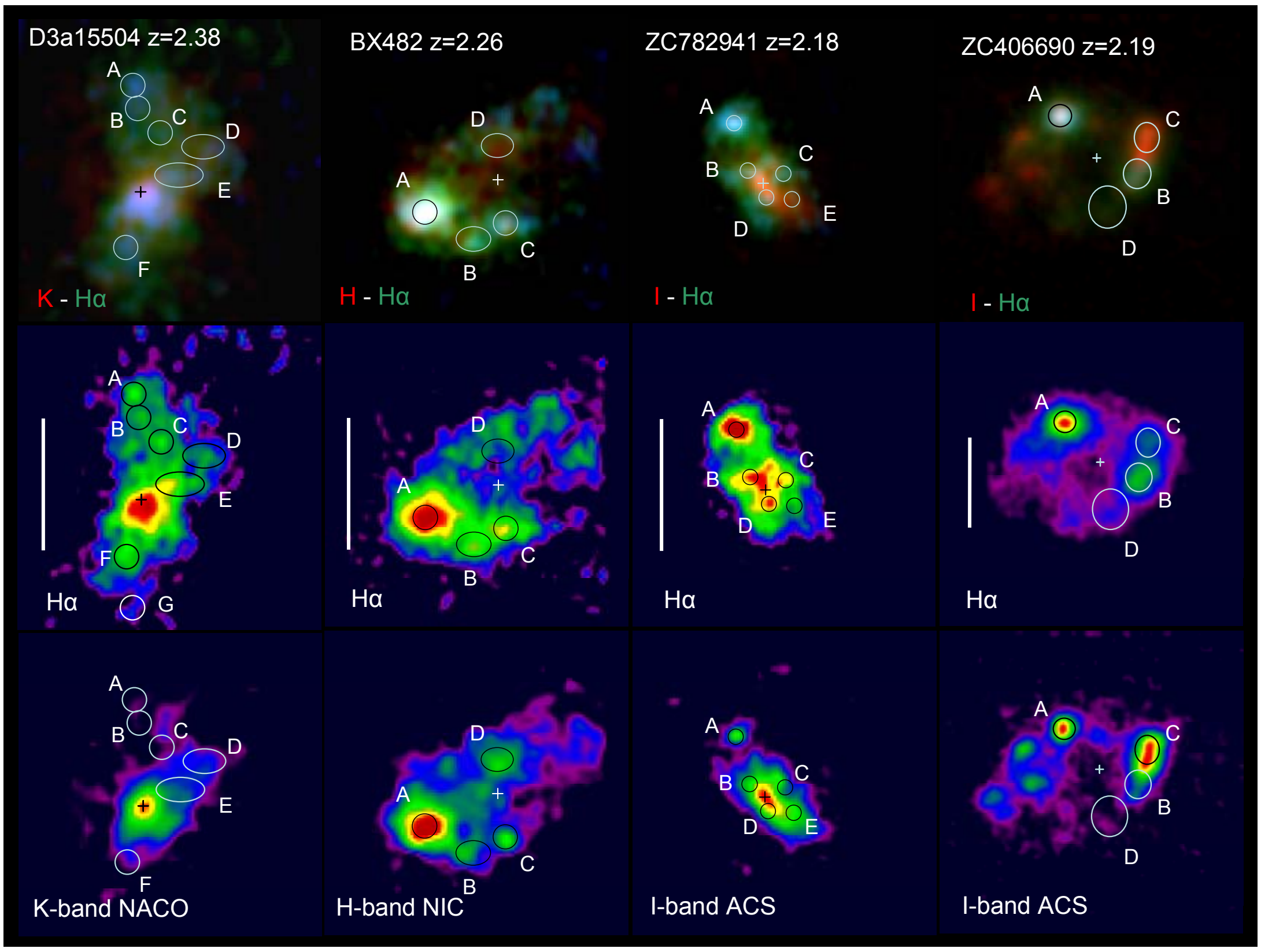

Figure 2. FWHM ~0.2” Hα and restframe UV/optical continuum images of four massive luminous z~2 SFGs. All maps have been re-binned to 0.025” pixels. Top row.: 3-color composites of integrated Hα line emission (red), and continuum (blue-green) images, along with the most prominent clumps identified by labels A, B…... Middle: Integrated SINFONI Hα emission. All four images are on the same angular scale, with the white vertical bar marking 1” (~8.4 kpc). Bottom. HST NIC-H band, ACS I-band, or NACO-VLT AO $K_s$-band images of the program galaxies, at about the same resolution as the SINFONI Hα maps**.** The color scale is linear and auto-scaled.

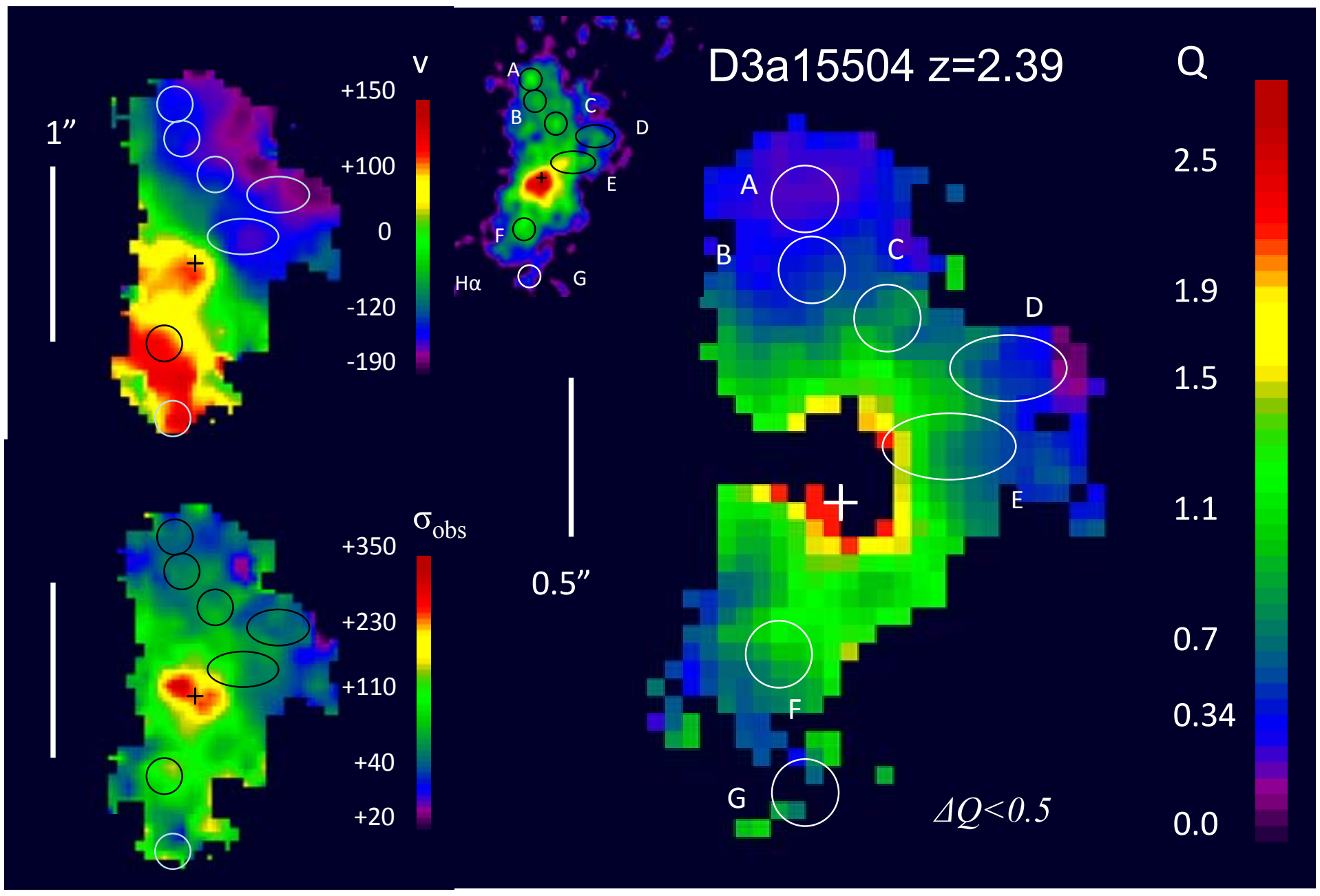

Figure 3. Hα Gaussian fit velocities (top left), Hα Gaussian fit dispersion (bottom left) and inferred Toomre *Q*-parameter (right, equation 2) for D3a15504. Shown in the top center is also the map of Hα integrated flux from Figure 2. The locations of the main clumps (Figure 1) found in the individual velocity channel maps are denoted by circles/ellipses. The Hα, velocity and velocity dispersion maps (resolution 0.18" FWHM) were re-binned to 0.025" pixels. For construction of the *Q*-map the data were smoothed to 0.25" FWHM. The typical uncertainties in the *Q*-values are ±0.05 to ±0.3 (1σ) throughout most of the disk of D3a15504. Pixels with $\Delta Q \geq 0.5$ are masked out.

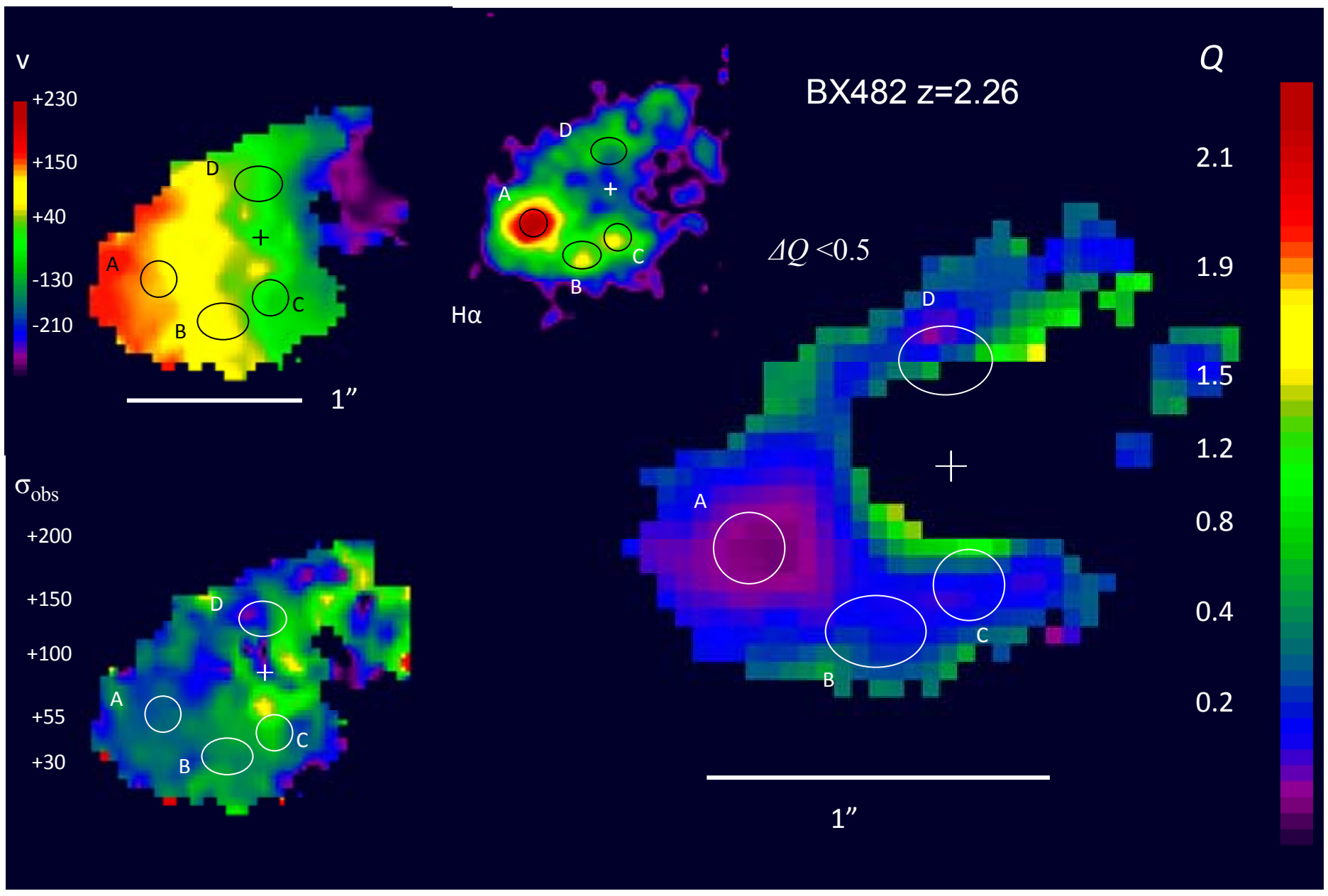

Figure 4. Maps of Hα Gaussian fit velocities (top left), Hα Gaussian fit dispersion (bottom left) and of the Toomre $Q$-parameter (right, equation 2) for BX482. Shown in the center is also the map of Hα integrated flux from Figure 2. The locations of the main clumps (Figure 2) are denoted by circles/ellipses. The Hα, velocity and velocity dispersion maps (resolution 0.18" FWHM) were re-binned to 0.025" pixels. For construction of the $Q$-map the data were smoothed to 0.25" FWHM. The typical uncertainties in the $Q$-values are ±0.03 to ±0.2 (1σ) along the bright ring of BX482. All pixels with $\Delta Q > 0.5$ were masked out.

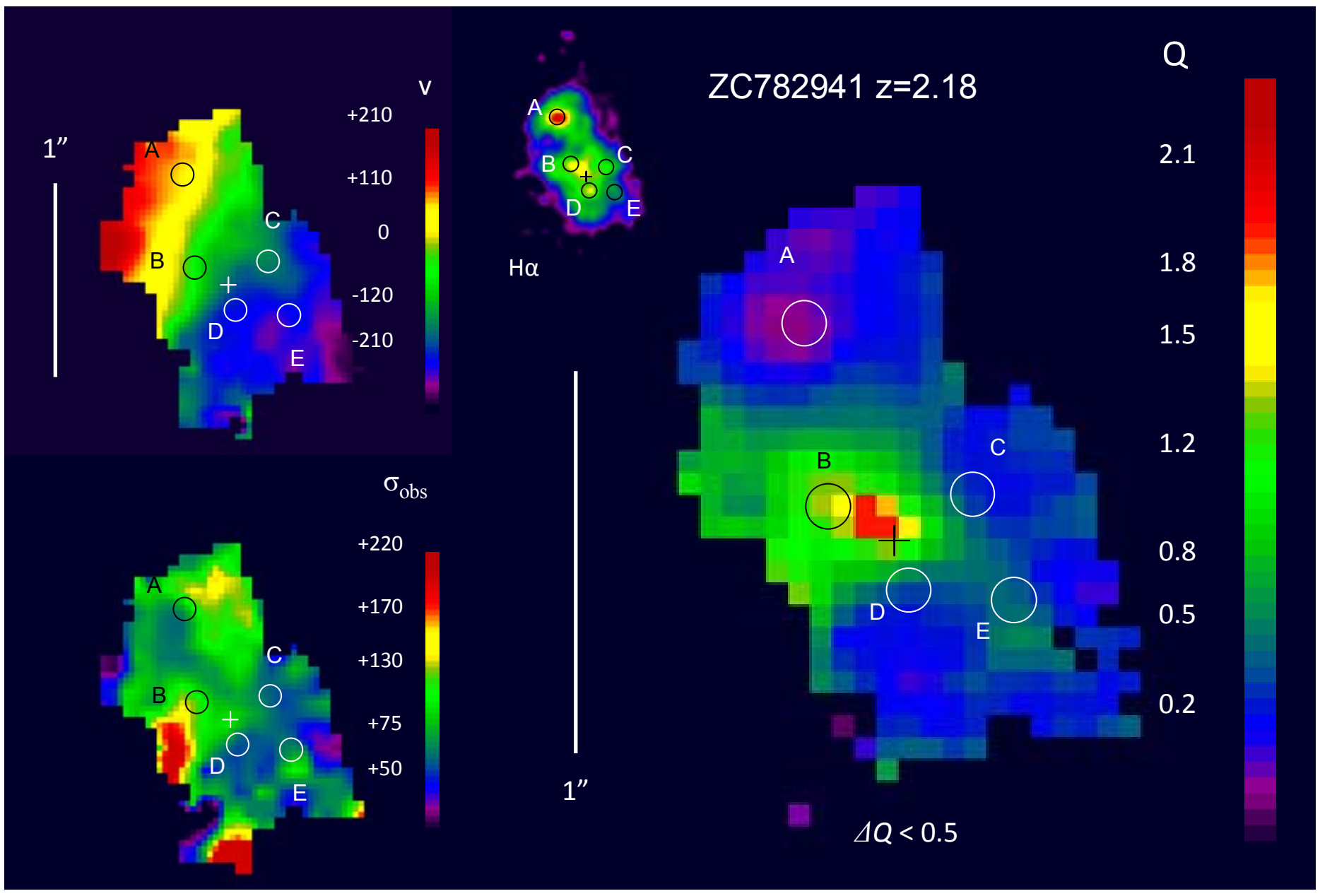

Figure 5. Maps of Hα Gaussian fit velocities (top left), Hα Gaussian fit dispersion (bottom left) and of the Toomre $Q$-parameter (right, equation 2) for ZC782941. Shown in the center is also the map of Hα integrated flux from Figure 2. The locations of the main clumps (Figure 2) are denoted by circles/ellipses. The Hα, velocity and velocity dispersion maps (resolution 0.18" FWHM) were re-binned to 0.025" pixels. For construction of the $Q$-map the data were smoothed to 0.25" FWHM. The typical uncertainties in the $Q$-values are ±0.06 to ±0.4 (1σ) for most of the outer disk of ZC782941. Pixels with $\Delta Q$ >0.5 were masked out.

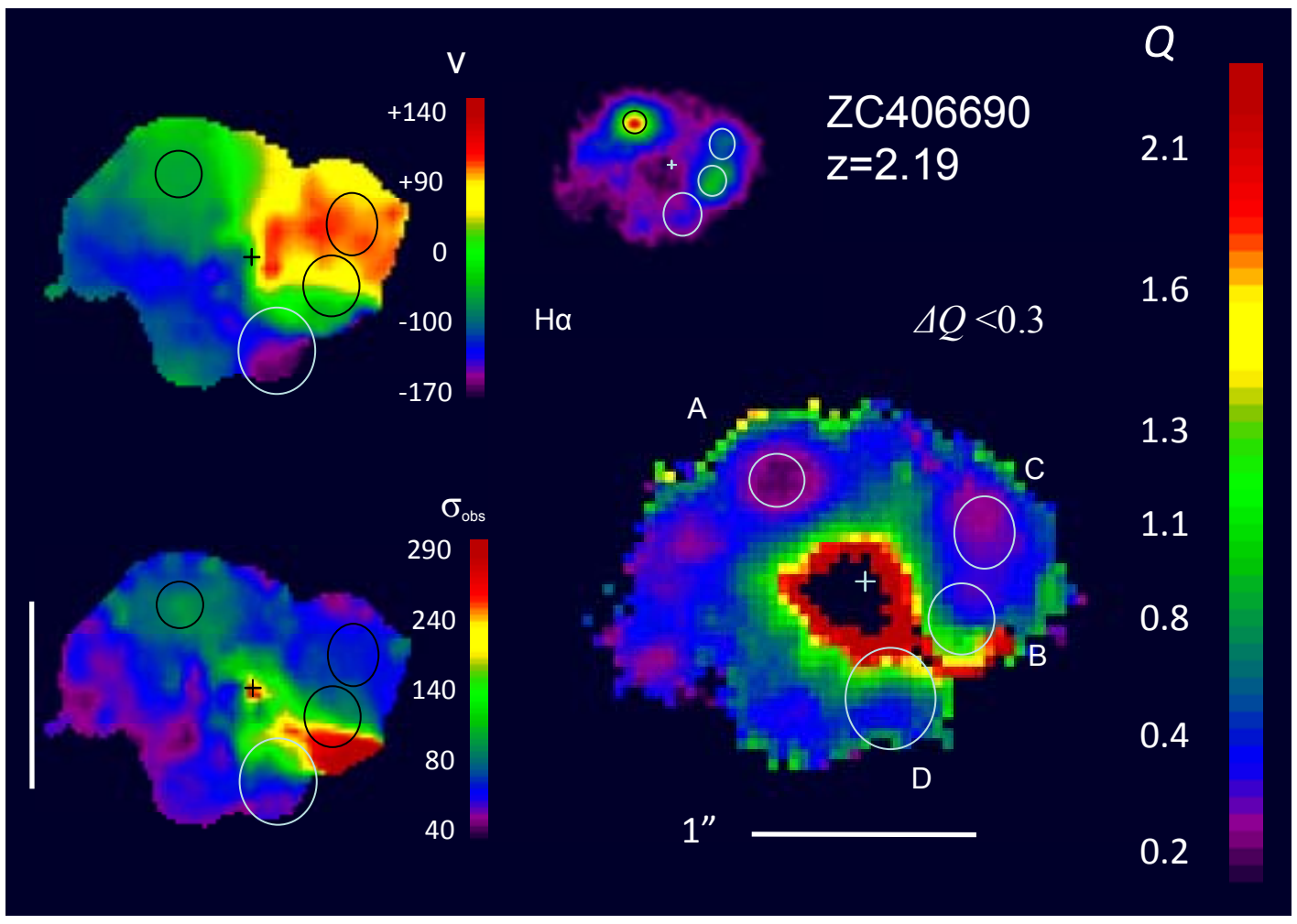

Figure 6. Maps of Hα Gaussian fit velocities (top left), Hα Gaussian fit dispersion (bottom left) and of the Toomre *Q*-parameter (right, equation 2) for ZC406690. Shown in the center is also the map of Hα integrated flux from Figure 2. The locations of the main clumps (Figure 2) are denoted by circles/ellipses. The Hα, velocity and velocity dispersion maps (resolution 0.22" FWHM) were re-binned to 0.025" pixels. The typical uncertainties in the *Q*-values are ±0.01 to ±0.1 (1σ) for most of the outer disk of ZC782941. Pixels with $\Delta Q$ >0.3 were masked out.

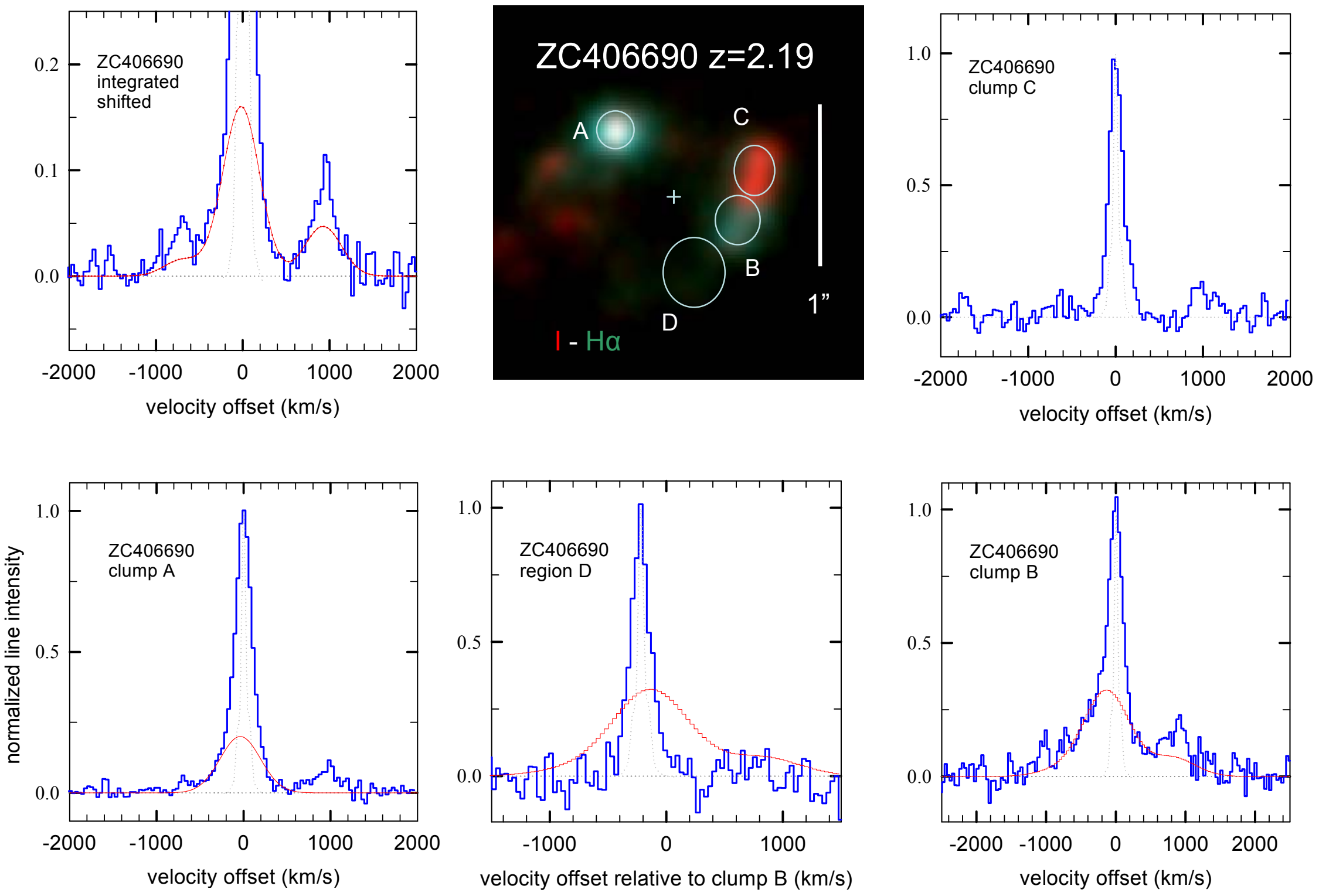

Figure 7. Hα spectra of ZC406690 (z=2.195), for the entire galaxy (top left), as well as for selected clumps marked in the central Hα-ACS I-band composite (Figure 2). The dotted grey curves in all panels denote the SINFONI spectral response profile measured from OH sky lines. Before construction of the integrated spectrum, the galaxy rotation was removed for each pixel. The thin red curves show the broad Hα (clump A) or broad Hα+[NII] (integrated spectrum, clump B) components obtained from multi-component Gaussian fits. For region D the thin red curve is the scaled broad Hα component of clump B, and the velocity scale is relative to the systemic velocity of nearby clump B.

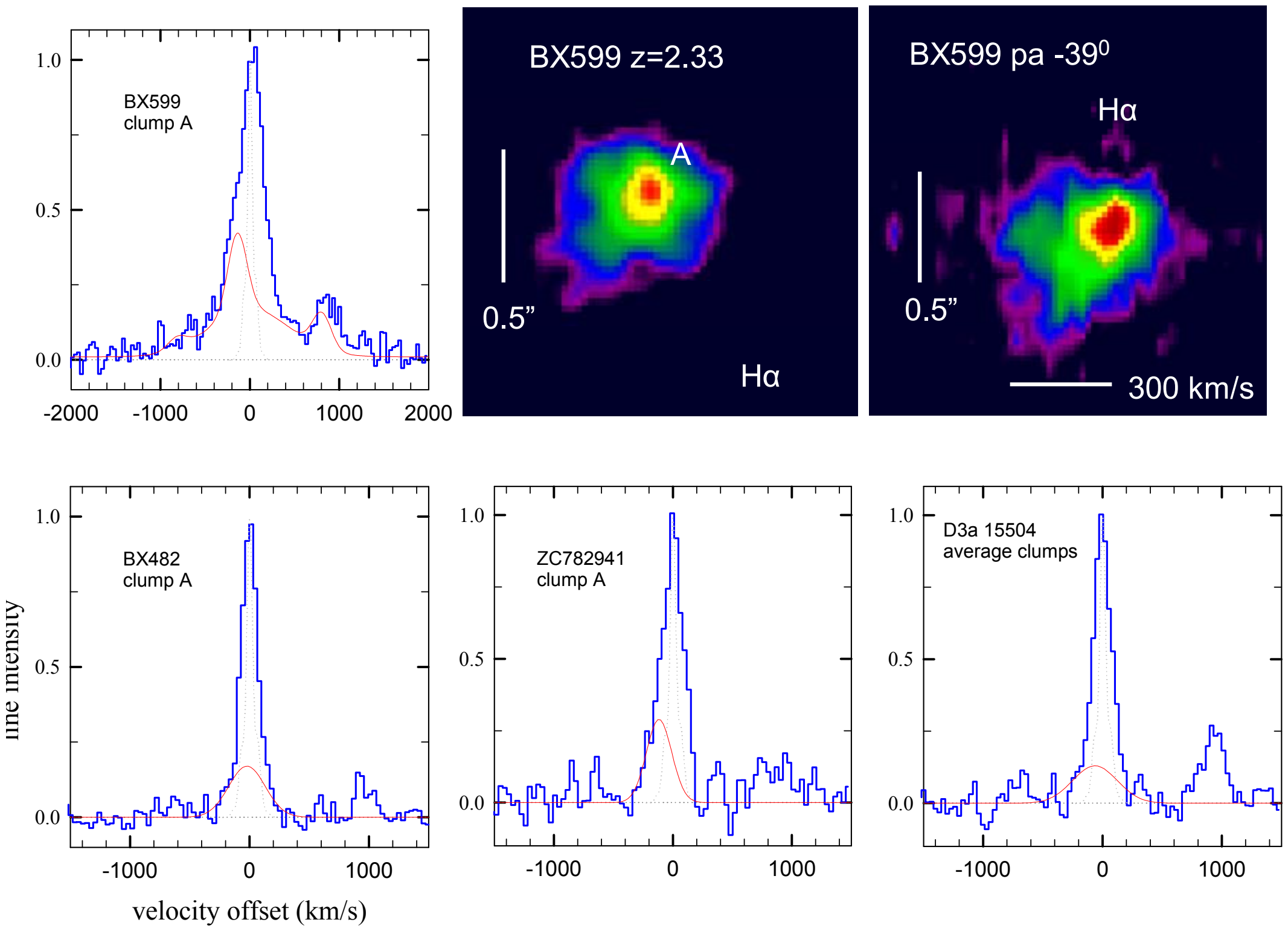

Figure 8. Spectra and images of BX599, BX482, ZC782941 and D3a15504. Top left: Hα spectrum toward the center of the compact dispersion dominated galaxy Q1623-BX599. Top center: integrated Hα map of BX599, sampled to 0.025" per pixel and smoothed to a resolution of 0.2" FWHM. Top right: position (up-down)-velocity (left-right) diagram of the Hα/[NII] emission of BX599 in a software slit at position angle -$39^0$ through clump A in the central panel, constructed from ~0.2" FWHM LGSF data, re-sampled to 0.025" and 16.7 km/s per pixel, half the original pixel scales. Bottom: Hα spectra of the brightest clump A in BX482 (left), clump A in ZC782941 (middle) and an average of clumps A-E in D3a15504 (right). The thin red curves denote fits of the broad Hα component, or the broad Hα+[NII] components in these clumps, as obtained from multi-component Gaussian fits. In the case of BX599 the broad Hα emission required two components. The dotted grey curve is the SINFONI spectral response function.

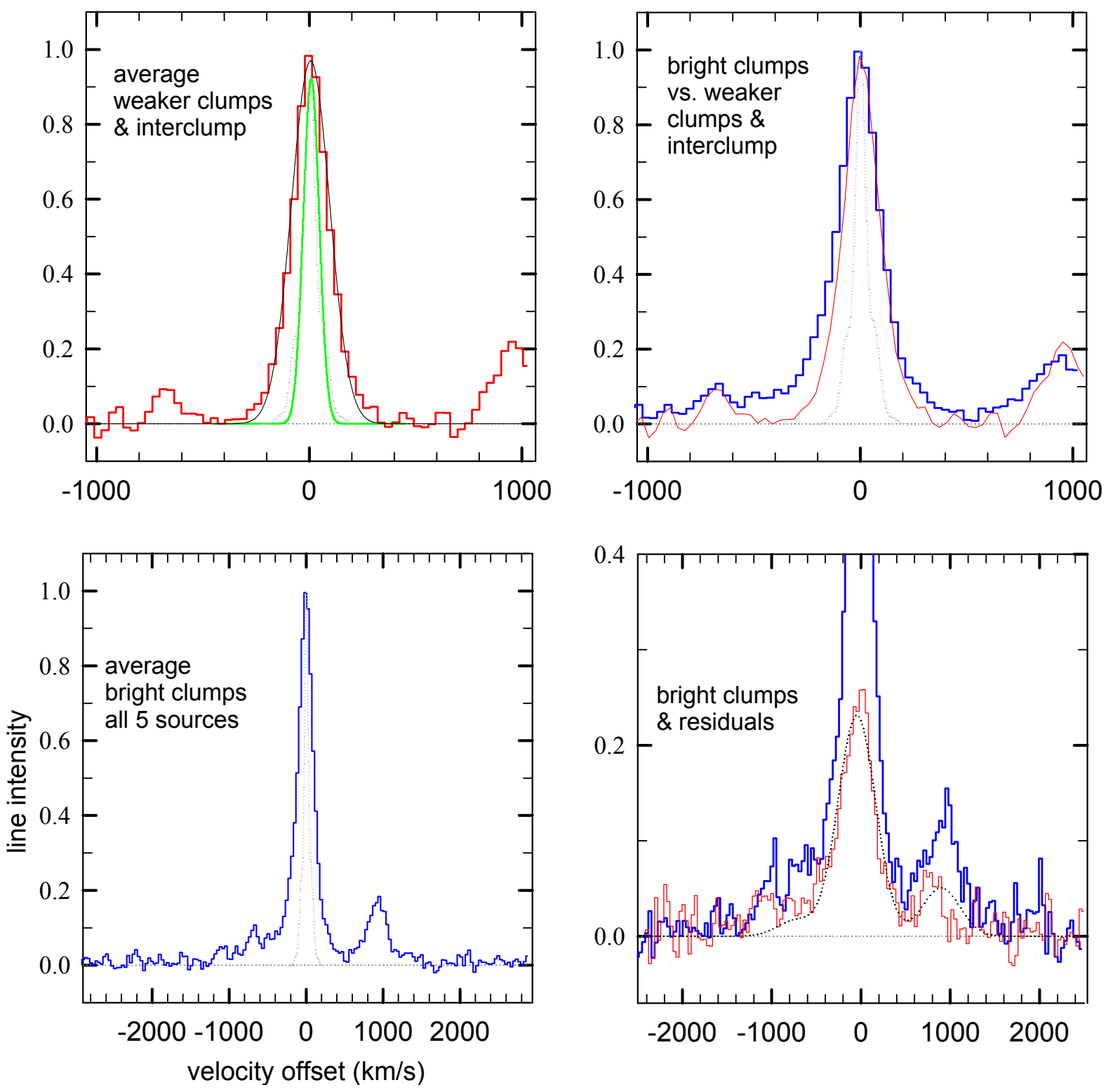

Figure 9. Average line profiles, compared to the instrumental line profile obtained from OH sky lines (dotted grey). Bottom left: Average Hα/[NII] spectrum (blue continuous) of the brightest clumps in D3a15504 (clumps A-F), BX482 (clump A), ZC782941 (clump A), ZC406690 (clumps A, B) and BX599 (clump A), after shifting to a common centroid. Top left: Average of 'interclump-weaker-clump-emission' in BX482, D3a15504 and ZC782941 (red continuous curve). The green continuous curve is a Gaussian fit (FWHM 86 km/s) to the instrumental profile, which is a good fit with some excess emission in the wings of the instrumental profile. The black continuous curve is the best fitting Gaussian to the 'interclump' emission profile

(FWHM 212 km/s, or σ=90 km/s). A fit with the instrumental profile gives a similar result (FWHM 202 km/s). Top right: Comparison of the interclump (red) and bright clump average (blue), demonstrating that the core of the bright clump emission is well fit by the interclump profile and that there is clear excess in the blue wing. Bottom right: Subtraction of 0.75 times the interclump profile from the bright clump profile (blue continuous) then results in the red residual profile, which is reasonably well fit by a Gaussian of FWHM 500 km/s centered at -48 km/s (black dotted).

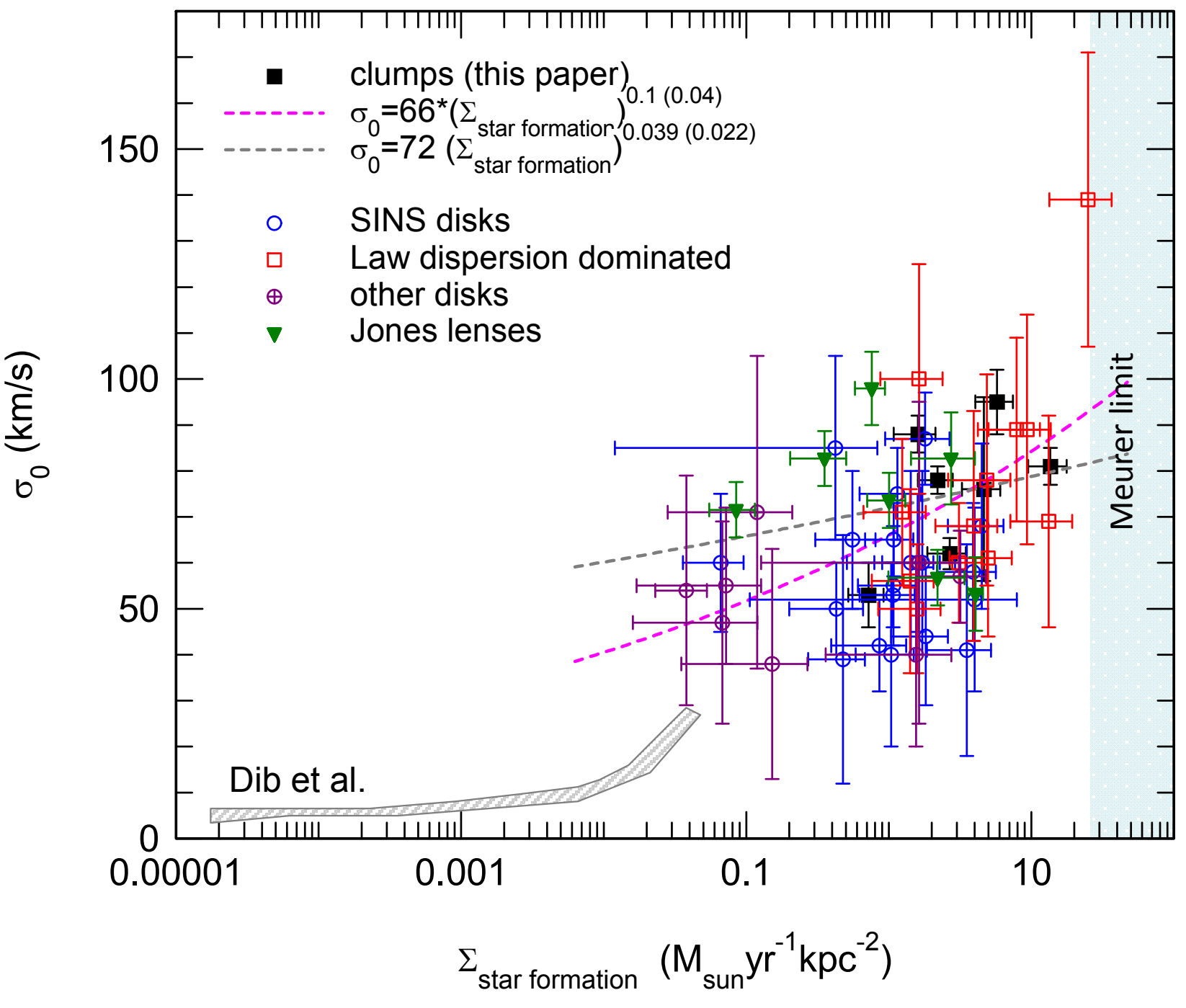

Figure 10. Dependence of intrinsic velocity dispersion on star formation surface density, for galaxy averages, as well as for individual giant star forming clumps. In all cases the effects of large scale velocity gradients, beam smearing and instrumental resolution were removed, if applicable, prior to estimating the local velocity

dispersion. Open blue circles (and 1σ uncertainties) denote galaxy averages in SINS z~1.5-2.5 disks (Förster Schreiber et al. 2006, 2009, Genzel et al. 2006, 2008, Cresci et al. 2009, this paper) and open purple circles denote galaxy averaged z~1-2 disks from Wright et al. (2007), van Starkenburg et al. (2008), Epinat et al. (2009) and Lemoine-Busserolle & Lamareille (2010). Open red squares are flux weighted galaxy averages of dispersion dominated z~1.5-2.5 SFGs from Law et al. (2009). Filled black squares denote the brightest clumps in BX482, D3a15504, ZC400690 and ZC782941, as well as the central region (clump) of BX599, and the filled green triangles mark the flux weighted dispersions in low-mass lensed z~2-3 LBGs of Jones et al. (2010). The dotted grey and dashed magenta lines are the best weighted and unweighted linear fits to the log-log representation of all data. The light blue shaded area on the right marks the region with $\Sigma \geq 26\ M_{\odot} yr^{-1} kpc^{-2}$, where no UV-bright star forming galaxies have so far been detected (Meurer et al. 1997, corrected to a Chabrier IMF). The grey hatched region shows the dependence of HI velocity dispersions in z~0 SFGs, as collected by Dib, Bell & Burkert (2006).

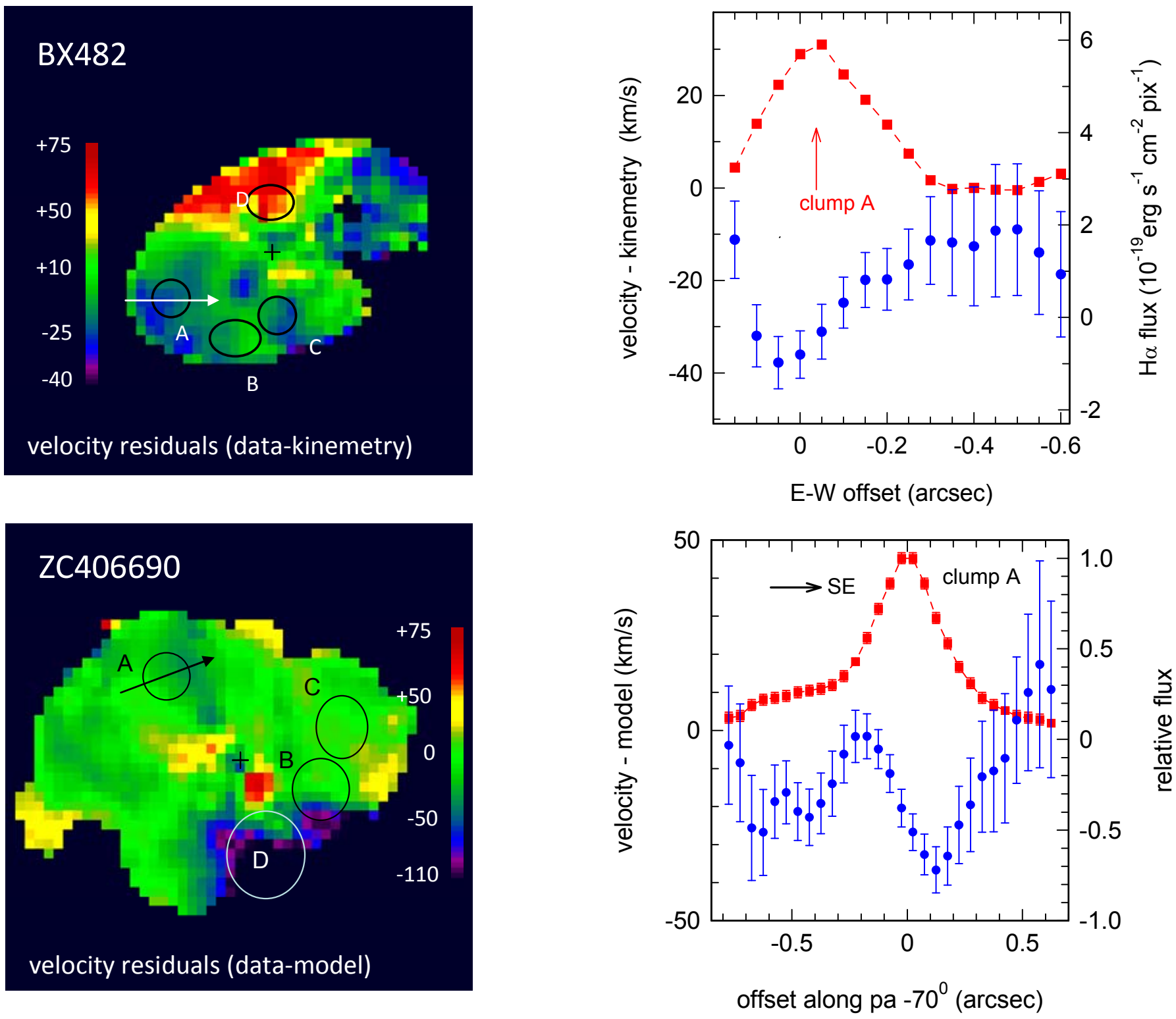

Figure 11. Velocity residual maps and position – velocity residual cuts across the brightest clump A in BX482 (top) and ZC406690 (bottom). The left panels are the residual maps (velocity (data) minus velocity (model or kinemetry)), the right panels give position- velocity residual (and ±1 σ errors) cross-cuts across the brightest clumps in each galaxy, along the direction of the galaxy's maximum velocity gradient (line of nodes). The red points and dashed curve denote the Hα flux (right vertical axis) and the blue points and continuous curve denote the residual velocity (left vertical axis).

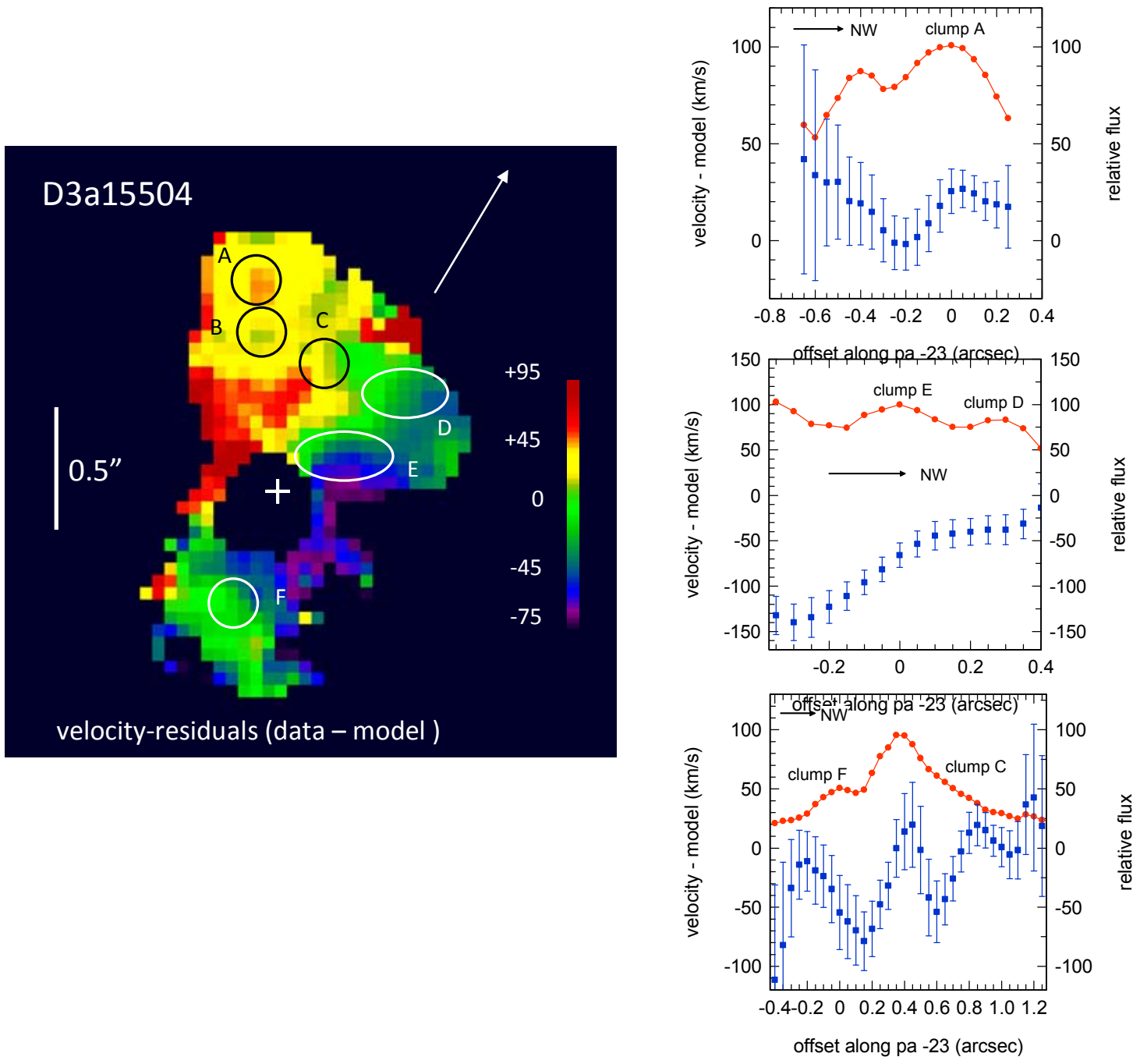

Figure 12. Velocity residual map of D3a15504 (left) and position-velocity cuts (right) through the five selected clumps along the line of nodes of the rotation of the galaxy (white arrow in left panel). The red points and dashed curves in the right panel denote the Hα fluxes (right vertical axis) and the blue points and continuous curves denote the residual velocities (left vertical axis).

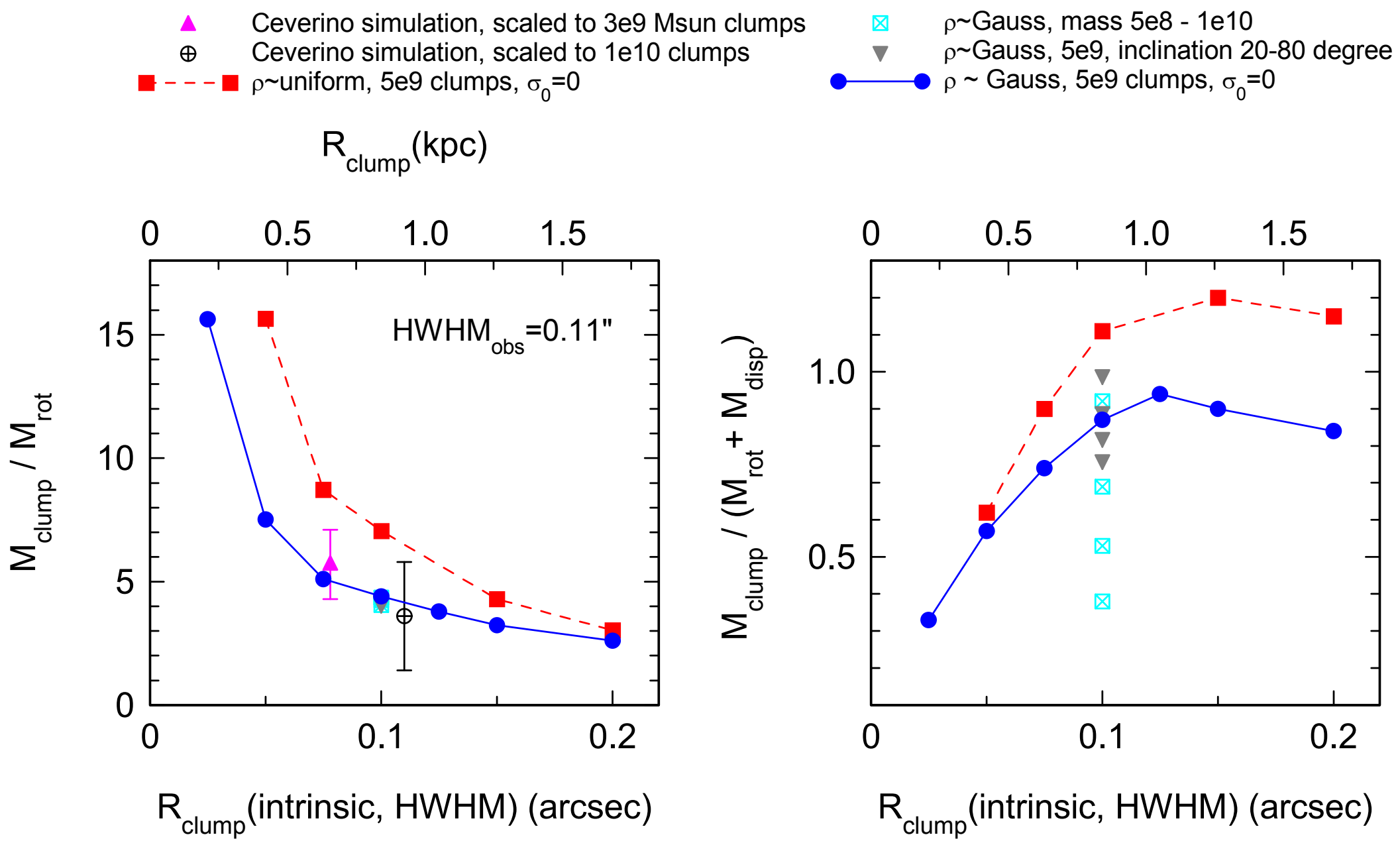

Figure 13. Properties of rotationally supported model clumps. Left panel: ratio of clump mass to the 'rotation' mass estimator in equation 6 (with b=1), for Gaussian and uniform density clumps of different intrinsic clump radii (HWHM for Gaussian, outer radius for uniform density), masses, and inclinations, convolved with the SINFONI spatial (FWHM 0.22") and spectral (FWHM 85 km/s) resolutions and sampled with 0.05" and 34 km/s pixels. Filled symbols denote the estimator in equation (6) applied at the observed clump HWHM. The crossed black circle and the filled magenta triangle denote the ratio derived from an average of four prominent clumps in a z=2.3 galaxy from the AMR simulations of Ceverino et al. (2010), but scaled to clump masses of $10^{10}$ and $3x10^9$ $M_\odot$, respectively. The simulated data were analyzed with the same method and effective resolution as the real SINFONI data

sets. The right panel gives the ratio of the clump mass to the mass estimator in equation (7), which combines the observed velocity gradient and velocity dispersion.

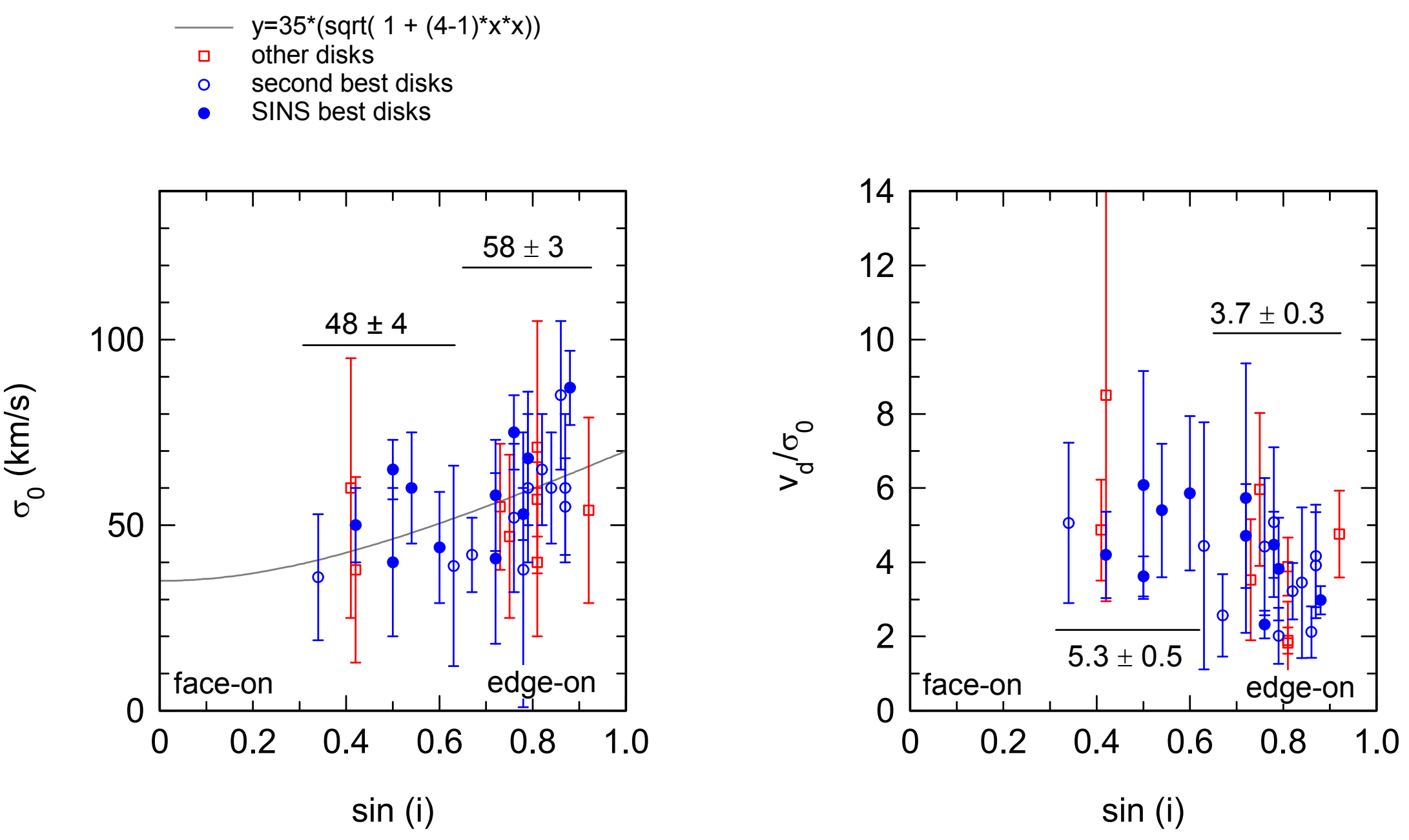

Figure 14. Dependence of intrinsic velocity dispersion $\sigma_0$ (left panel, corrected for large scale velocity gradients, beam smearing and instrumental resolution) and $v_c/\sigma_0$ (right panel) on the inferred inclination (typical uncertainty ±0.1 in *sin(i)*) for the SINS disks (filled and open blue circles) and disks from other publications (open red squares, see references in the caption of Figure 10). Mean and 1σ uncertainty in the mean are indicated for the ranges *sin(i)* less than and greater than 0.64.

# Appendix A

We have explored possible intra-galaxy variations of the intrinsic velocity dispersion $\sigma_0$ as a function of position in the deep AO-data on the four SFGs reported in this paper. To remove large scale velocity dispersion gradients artificially created by beam-smeared velocity gradients (caused by rotation, for instance) we analyze residual velocity maps (section 2.2). Figures A1 and A2 give the results. We have found that the key issue is error estimation and significance. Overlays of spatially resolved residual velocity dispersion as a function of position on the Hα maps (Figures A1 & A2) at first glance seem to show in several galaxies a mild trend for Hα bright clumps to be associated with lower velocity dispersion than in the surrounding regions. However, a more detailed analysis of the fit uncertainties derived from our Monte Carlo bootstrapping method reveals that most of this trend is due to poor signal to noise ratio in the interclump region. When weighted fits to the pixel-to-pixel variations of the residual velocity dispersions $\delta\sigma$ (section 2.2) as a function of $\Sigma_{H\alpha}$ are considered for the higher signal to noise ratio regions, only relatively weak, or no (positive) correlations are found. These mild trends are in good agreement with the galaxy averages and selected clumps shown in Figure 10.

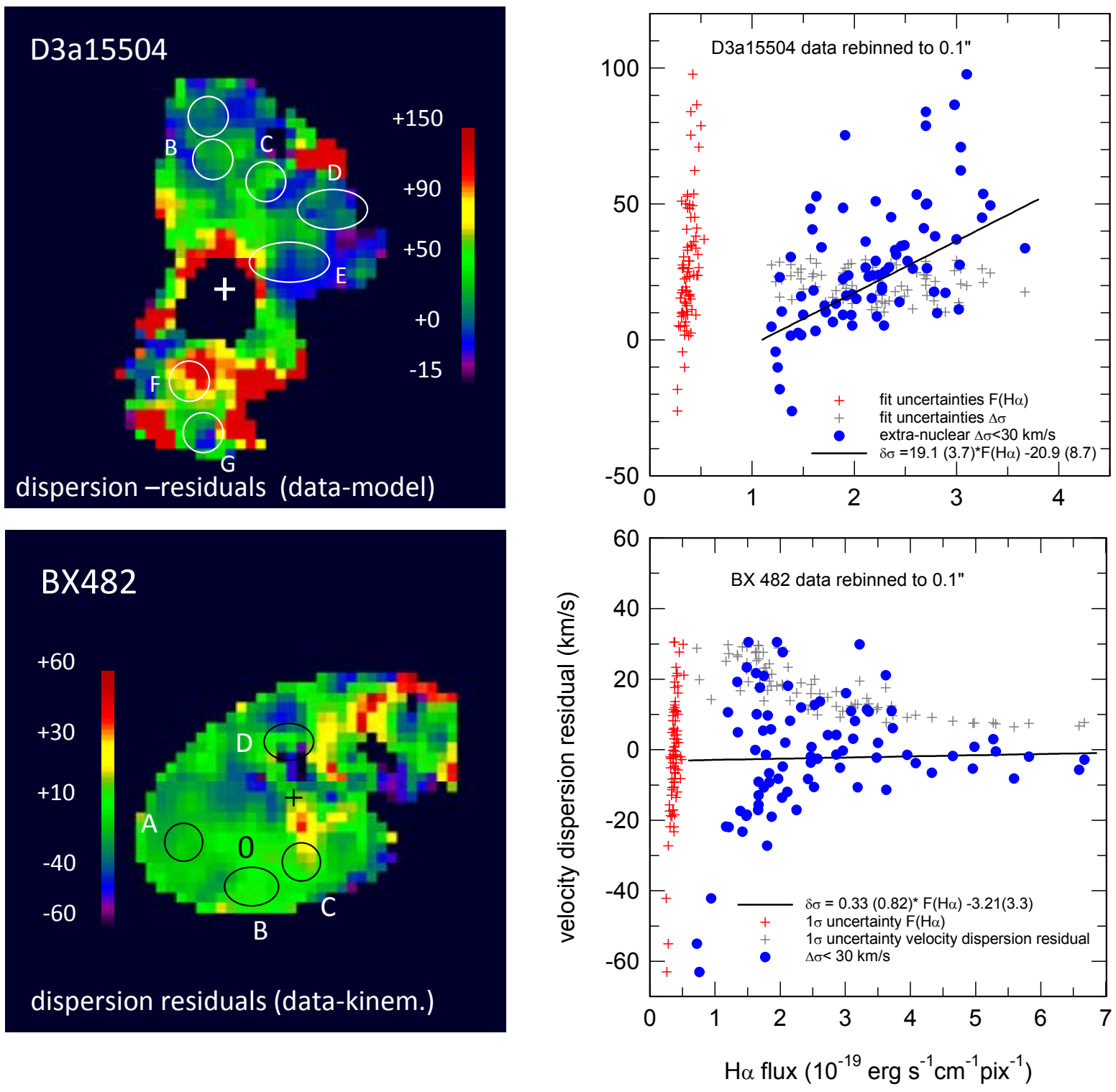

Figure A1. Dependence of residual velocity dispersion on Hα surface brightness for D3a15504 (top) and BX 482 (bottom). The left maps show the spatial distribution of the residuals (δσ= σ(data) minus σ(model) in upper plots and σ(data) minus σ(kinemetry averages) in lower plots) within each of the two galaxies. The input velocity dispersions are intrinsic values after removal of the instrumental broadening and beam smeared large scale velocity gradients (such as rotation). Minima and maxima of the color codes are indicated, as well as the location of the prominent clumps from Figure 2. Analysis of the 2d-maps requires careful attention to error analysis, as most of the obvious variations in the D3a15504 velocity dispersion residual map, for instance, is caused by biases due to large uncertainties of σ in low surface brightness regions. The right maps show the pixel to pixel correlations, after culling low significance data. Filled blue circles denote those data with fit uncertainties Δσ less than 30 km/s, after rebinning the data to 0.1” per pixel. For these data the red and grey crosses denote the distribution of Hα surface brightness and δσ 1 σ errors, respectively. The black line is the weighted ($w_i=1/\Delta\sigma_i^2$) linear regression fit to the filled blue circles. Fit parameters (and 1σ uncertainties in parentheses) are given in the legend.

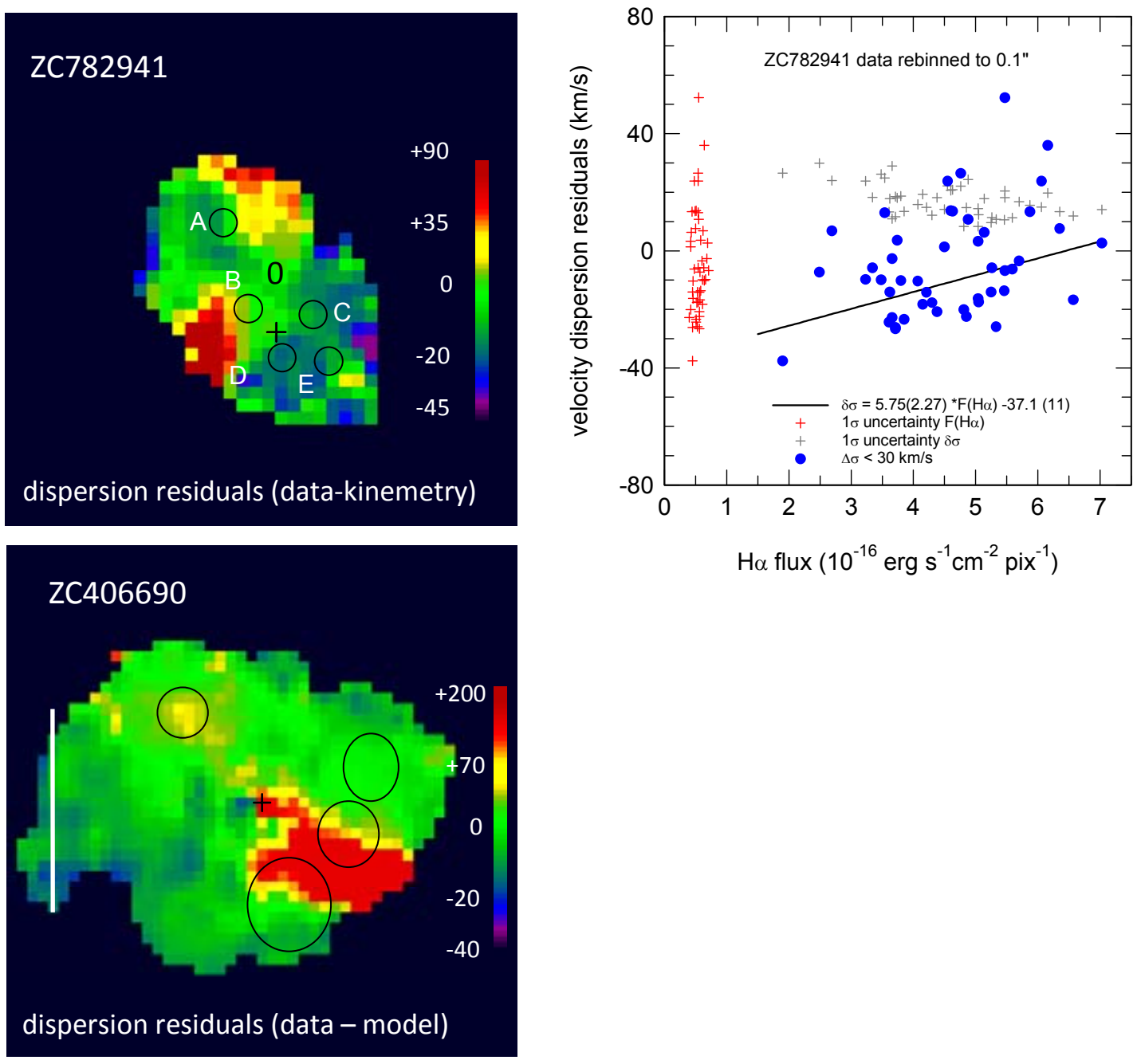

Figure A2. Dependence on residual velocity dispersion on Hα surface brightness for ZC782941 (top) and ZC406690 (bottom). The left maps show the spatial distribution of the residuals (δσ= σ(data) minus σ(model) in lower plots and σ(data) minus σ(kinemetry averages) in upper plots) within each of the two galaxies. Minima and maxima of the color codes are indicated, as well as the location of the prominent clumps from Figure 2. The input velocity dispersions are intrinsic values after

removal of the instrumental broadening and beam smeared large scale velocity gradients (such as rotation). Analysis of the 2d-maps requires careful attention to error analysis, as the apparent increases in the velocity dispersion residual map at the southern and northern edges of ZC782941, for instance, are caused by biases due to large uncertainties of σ in low surface brightness regions. In the case of the ZC406690 dispersion residual map, the dispersion enhancements in the integrated line profiles (as shown) near/on clumps A and B are almost entirely caused by the broad wind component affecting the rms dispersion of the overall line profiles (Figure 7). The narrow component does not show larger dispersions toward these clumps. The right graph shows the pixel to pixel correlation for ZC782941, after culling low significance data. Filled blue circles denote those data with fit uncertainties Δσ less than 30 km/s after re-binning the data to 0.1” per pixel. For these data the red and grey crosses denote the distribution of Hα surface brightness and δσ 1 σ errors, respectively. The black line is the weighted ($w_i=1/\Delta\sigma_i^2$) linear regression fit to the filled blue circles. Fit parameters (and 1σ uncertainties in parentheses) are given in the legend.

# Appendix B

In this section we discuss our quantitative analysis of the broad Hα emission for deriving estimates of the mass outflows rates in the various galaxies of our sample.

## B.1 Outflow estimates for photoionized, recombining gas

We estimate masses and outflow rates of the warm ionized gas in our program galaxies/clumps for three simple models. In all models we assume an outflow into solid angle $\Omega$ and a radially constant mass loss rate and outflow velocity. The latter is motivated by observations of the M82 outflow (e.g. McKeith et al. 1995) and theoretical work on both energy and momentum driven outflows (Veilleux et al. 2005, Murray et al. 2005). In the first two models we assume that the gas is photoionized, and in case B recombination with an electron temperature of $T_4=10^4$ K (Osterbrock 1989). In the first model the *average* electron density scales with $R^{-2}$ (for a constant mass outflow rate) but the *local* electron density of filaments or compact clouds from which the Hα emission derives does not vary significantly with radius and takes on a value of $< n_e^2 >^{1/2} \sim 100$ cm$^{-3}$. This choice is motivated by observations of electron densities in the z~0 outflows derived from the [SII] 6718/6733Å doublet (Heckman, Armus and Miley 1993, Lehnert & Heckman 1996). In the second model we assume that the ionized gas fills the entire volume of the outflow cone. In that case both the average and the local electron density scale with $R^{-2}$. For purely photoionized gas of electron temperature $T_4 = 10^4 K$ and case B recombination, the effective volume emissivity is $\gamma_{H\alpha}(T) = 3.56 \times 10^{-25} T_4^{-0.91}$ erg cm$^{-3}$s$^{-1}$ (Osterbrock 1989). The total ionized mass outflow rate can then be obtained from the extinction corrected Hα luminosity $L_{H\alpha,0}$ via

$$
\begin{aligned}
L_{H\alpha,0} &= \gamma_{H\alpha}(T) \int \Omega R^2 n_e(R) n_p(R) dR, \\
M_{HII,He} &= \mu \cdot \int \Omega R^2 n_p dR = \frac{\mu L_{H\alpha,0}}{\gamma_{H\alpha}(T) n_{eff}}, \text{ and} \\
\dot{M}_{out} &= \Omega R^2 \mu \, n(R) v_{ex} = \zeta \cdot M_{HII,He} \cdot \frac{v_{ex}}{R_{out}} \qquad \text{(B1).}
\end{aligned}
$$

Here $\mu = 1.36 \cdot m_p$ is the effective mass, for a 10% helium fraction, and $v_{ex} \approx \Delta v_{\max}$ in Table 2. $R_{out}$ is the outer radius of the outflow that initially is launched near the center of the disk/clump at $R_{in} << R_{out}$. To compute an extinction corrected luminosity for the broad component in Table 2 we corrected the observed fluxes both for the general galaxy obscuration estimated from the UV-continuum colors (and a Calzetti 2001 extinction law with $A_{gas}=A_{stars}/0.44$), as well as for additional differential extinction through the clump/galaxy, as estimated very approximately from the asymmetry of the broad line emission (factor $\gamma_{red}$ in Table 2). The expressions for $\zeta$ and $n_{eff}$ depend on the assumed geometry and density distribution of the outflow and are different for our two models. For model 1 we adopt $n_{eff} = < n_e^2 >^{1/2} = 100$ cm$^{-3}$ and $R_{out} \sim R_{HWHM}$. For model 2 we have $n_{eff} = n_{in}(R_{in}) R_{in} / R_{out}$ and given the galaxy wide distribution of the outflowing gas in ZC406690 we assume $R_{out} \sim 6$ kpc$\sim R_{disk} \sim 10\ R_{in}$. This assumption is motivated by the modeling of UV absorption line gas velocities $v_{out}(R)$ (Figures 23 and 24 in Steidel et al. 2010). Because of the $n_e^2$-scaling of the Hα emission, $R_{HWHM} \sim 2.3\ R_{in}$. These choices for $R_{out}$ probably bound the true (emission weighted value) of $R_{out}$ from below and above. For both models (constant expansion velocity) $\zeta=1$.

For model 2 we assume that near the launch point the ionized gas in the wind is in pressure equilibrium with the star forming gas in the clump, such that $n_{in}(R_{in}) \sim n_{e,clump}$. One direct estimate for this base density comes from the [SII] 6718/6733Å ratio that can be determined empirically for several of the galaxies/clumps in Table 2, as shown in Figure B.1. The average/median of these estimates (row 15 in Table 2) is 900 cm$^{-3}$. Unfortunately the uncertainty is very large for each of the individual entries, sometimes including zero or infinite density. A second estimate for the base density comes from estimates of the total mass/gas densities in the giant clumps. Assuming that the clumps are virialized their average matter densities (gas plus stars) are $<n_{clump}>\sim\frac{\sigma_0^2}{\mu G R_{clump}^2}$, corresponding to densities between 10 and 40 $M_{\odot}pc^{-3}$, or 300 to 1100 H $cm^{-3}$ for the clumps in Table 2, Figure 10 and in the well resolved lensed SFGs observed by Jones et al. (2010). Assuming gas fractions of about 50% (Tacconi et al. 2010, Daddi et al. 2010a), this corresponds to (cold) gas densities of 150 to 550 $cm^{-3}$. Another approach is to take the observed star formation densities and convert to gas densities with the Kennicuut-Schmidt relation, as already discussed in 3.1. For clump A in ZC406690, for instance, this approach yields an average (cold gas) density of ~70 $cm^{-3}$. The bottom line is that these *average* cold gas densities across the giant clumps are very similar to those in Milky Way Giant Molecular Clouds (GMCs, $<n_{gas,GMC}>$ ~ 170 $cm^{-3}$ (e.g. Blitz 1993, Harris & Pudritz 1994). However, in these GMCs the actual *star forming gas* has much higher densities, $n_{SF}\sim10^{4.5}$ $cm^{-3}$, with local pressures of $n_{SF}T_{SF}\sim10^{6.4\pm0.5}$ K $cm^{-3}$. Assuming again pressure equilibrium the ionized gas in these star forming regions would have electron densities of $10^{2.6\pm0.5}$ $cm^{-3}$ for $T_e$(MW)~7000 K. If the high-z clumps have similar conditions, this consideration yields an upper limit to $n_{in}(R_{in})\sim1200$ $cm^{-3}$, similar to the average value from the [SII] line ratios in Table 2. For model 2 we adopt $n_{in}(R_{in})\sim1000$ $cm^{-3}$. With these assumptions model 2 yields outflow rates 4 times lower than for model 1.

Rows 21 & 22 of Table 2 list the derived ionized gas outflow rates for these case B photoionization models, for individual clumps and galaxy averages in the five galaxies of our sample. The key finding in all clumps considered here is that in the photoionized case the derived outflow rates are comparable to or larger than the star formation rates, in one case (clump B in ZC406690) by a factor of 8.4 in the average of models 1 and 2.

## B.2 Outflow estimates for collisional ionization

If the gas temperature is high enough, it is possible that the gas responsible for the Hα emission is collisionally ionized in the wind. Based on [SII]/Hα, [OI]/Hα and [NII]/Hα line ratios Heckman, Armus and Miley (1990), Shopbell & Bland-Hawthorn (1998) and Veilleux and Rupke (2002) find that the outer region of the outflowing gas of M82 and NGC 1482 is likely shock-heated by the galactic wind fluid, while the inner region is photoionized by the starburst. Veilleux and Rupke (2002) characterize the ionization mechanism based on the [NII]/Hα ratio, such that values less than 0.5 are representative of HII regions and therefore indicative of ionization by O stars, whereas larger values characterize HH objects and other regions for which shocks are important. The low [NII]/Hα ratio found in the inner region of the M82 outflow is unusual for a galactic wind and suggests that the gas is ionized by a relatively young and active starburst (Veilleux and Rupke 2002). They further argue that, in the absence of an AGN, the large values of [NII]/Hα found in the outer regions of the M82 wind nebulae and characteristic of other galactic winds require an additional heating source, namely shock-heating by the wind fluid. If the gas is shock-heated, ionization can occur either through photoionization by EUV/soft X-rays produced in the shock, or through collisional processes (Heckman, Armus & Miley 1990).

We thus consider as our third model collisional ionization and excitation of line radiation, which will dominate at high densities and temperatures. A simple model estimating the photoionizaton rate based on the size of clump A in ZC406690, an electron density of 100 $cm^{-3}$, an average cross-section for ionization of $10^{-18}$ $cm^2$, and photoionization of gas located 3 kpc from the disk, suggests that collisional ionization begins to dominate at $T \geq 2 \times 10^4$K.

We estimate the mass outflow rate assuming collisional excitation and ionization at $T = 2\times10^4 K$, which is also the peak temperature for collisional-based emission (Goerdt et al. 2010). For $T = 10^5 K$, the outflow rates are about an order of magnitude larger. The Hα luminosity becomes

$$L_{H\alpha} = \gamma_{H\alpha}(T)\int \Omega R^2 n_e(R) n_p(R) dR + \ h\nu_{H\alpha} q_{exc}(T)\int \Omega R^2 n_e(R) n_H(R) dR \qquad \text{(B2)},$$

where $\frac{n_H}{n_p} = \frac{\gamma(H\alpha)}{h\nu_{H\alpha} q_{ion}}$, $n_p \approx n_e \approx 100\ cm^{-3}$ and $q_{exc} = 7.1\times10^{-12}$ $\mathrm{cm^3 s^{-1}}$ and $q_{ion} = 3.4\times10^{-12}$ $\mathrm{cm^3 s^{-1}}$ are the collisional excitation and ionization rate coefficients, at $T_e$~2x$10^4$ K from Osterbrock (1989). The mass outflow rates for collisional-excitation are listed in row 22 of Table 2. The bottom line is that if the outflowing gas comes from collisionally excited gas at $T_e$~2x$10^4$ K, the inferred masses and outflow rates are lowered by about a factor of 2 compared to the average case B recombination cases (row 23 in Table 2). If the temperature is lower, the difference is less.

We find relatively small values of [NII]/Hα for the broad components of clumps A and B in ZC406690 ([NII]/Hα ~0.3), as well as for the stacked spectrum in Figure 9 ([NII]/Hα ≤0.4). These ratios favor the photoionization model. However, the analysis of Veilleux and Rupke (2002) is based on near-solar metallicity galaxies in the local universe, whereas several of the high-z SFGs are less chemically evolved, and thus one might expect the critical [NII]/Hα -value differentiating between shock-heated and photoionized gas to be lower as well. Nevertheless it is likely that the gas outflowing from clumps A and B of ZC406690 is primarily photoionized and not shock-heated, based on the relatively low [NII]/Hα values in the broad component of emission from these clumps, and the fact that they appear to be very young, extremely active star forming regions. We thus will adopt the case B/photoionization outflow rates as our base values for the further analysis.

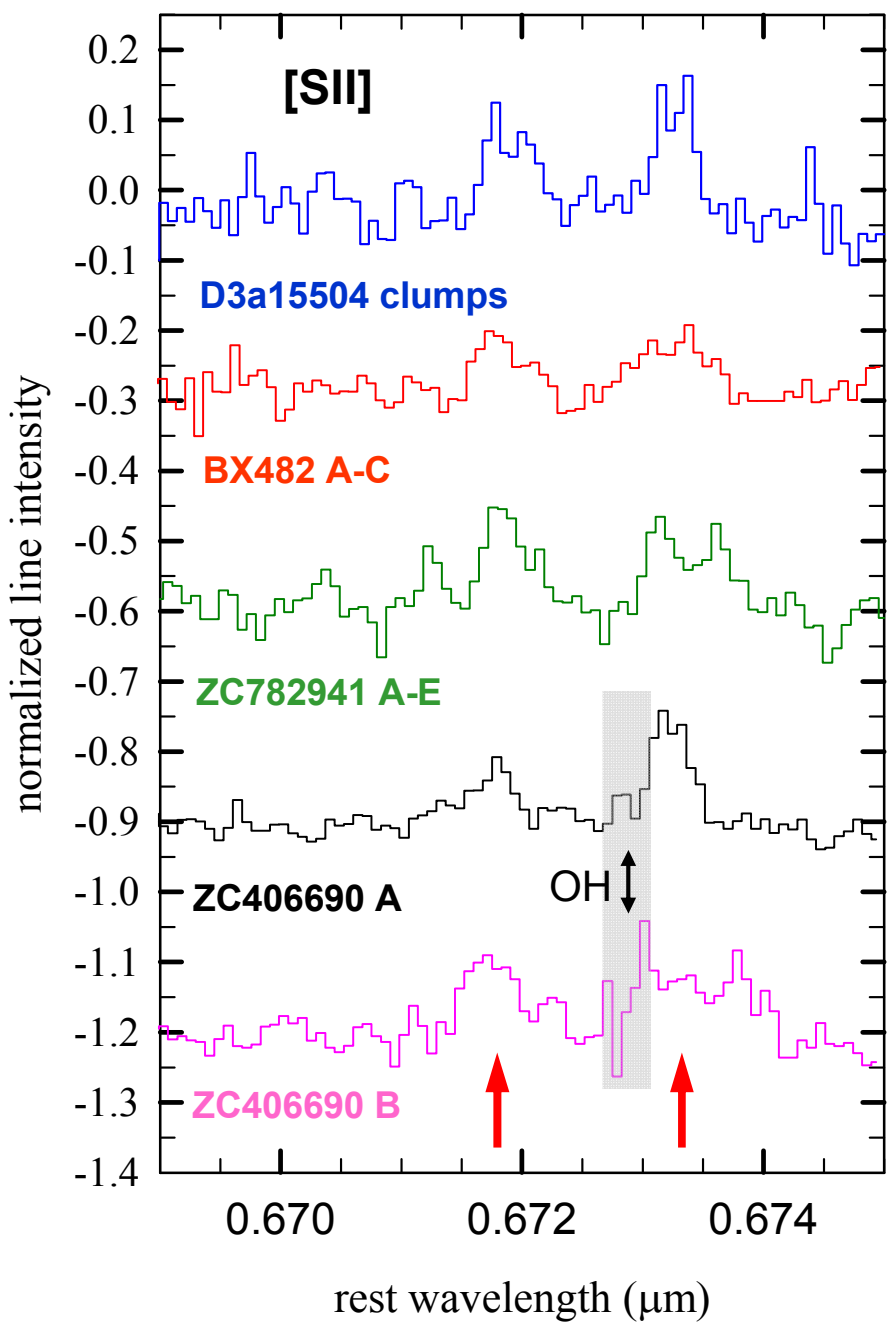

Figure B.1. [SII] 6718/6733Å doublet spectra for the most prominent clumps of four of our program galaxies. The locations of the two lines are shown by red arrows. An OH line strongly affects the blue side of the 6733 line in ZC406690 A/B.

# Appendix C

In this section we derive the metal enrichment timescales using the “closed” and “leaky” box models from Erb (2008). Erb’s models are derived from the Kennicutt-Schmidt-relation (Kennicutt 1998a) in order to determine the time dependence of star formation and the effect of infall and outflow on the metallicity. For the “closed box” model

$$\frac{d\left(ZM_g\right)}{dM_*} = y\alpha\left(1-Z\right)-\alpha Z, \text{ resulting in}$$

$$\frac{dZ}{dt} = \frac{y\alpha\left(1-Z\right)SFR}{M_g}, \text{ for } SFR = \frac{dM_*}{dt} \text{ and } M_g = M_i - \alpha M_* \quad \text{(C1)},$$

where $M_g$ is the gas mass, $M_i$ is the total initial mass, $M_*$ is the stellar mass, $\alpha$ is the mass fraction locked in stars (which we take as 0.8). The parameter $y$ is the true yield, which is the ratio of the mass of metals produced and ejected by star formation over the mass locked in stars. We follow Erb (2008) and adopt $y = 0.019$. Assuming that $y$, $\alpha$ and the star formation rate *SFR* are constant, and using the observed gas mass, as calculated with the Kennicutt-Schmidt relation, and our observed value of the metal mass fraction, $Z\left(t_f\right)$, we solve the differential equation numerically to determine $t_Z$ (the time required to produce $Z\left(t_Z\right)$) (see row 26 in Table 2).

We also consider a “leaky box” model (Erb 2008), where an outflow component is included with an outflow rate equal to that calculated for the photoionization model with a constant local density (Table 2, row 20). In this case, equation (C1) becomes

$$\frac{dZ}{dt} = \frac{y\alpha\left(1-Z\right)SFR}{M_g} \text{ for } M_g = M_i - \alpha M_* - f_0 M_* \quad \text{(C2)},$$

where $f_0 = \frac{\dot{M}_{out}}{SFR}$ is the mass fraction in the outflow with respect to the *SFR* and $\dot{M}_{out}$ is the mass outflow rate.

# Table 1. Observing Log

| Galaxy | band/pixel scale | mode | FWHM resolution (arcseconds) | integration time, observing date | reference |
|---|---|---|---|---|---|
| Q1623-BX599 (z=2.332) | K 0.05”x0.1” | LGS | 0.23” | 2h00<br>2010 Apr 12-13 | Erb et al. 2006b, Förster Schreiber et al. 2009 |
| Q2346-BX482 (z=2.258) | K 0.05”x0.1” | LGS | 0.25” | 9h30<br>2007 Oct 27-29<br>2007 Nov 13-15<br>2008 Jul 27-31<br>2009 Nov 11-13 &17 | Erb et al. 2006b Genzel et al. 2008, Cresci et al. 2009; Förster Schreiber et al. 2009 |
| D3a15504 (z=2.383) | K 0.05”x0.1” | LGS, NGS | 0.18” | 18h40<br>2006 Mar 16-20<br>2009 Apr 30<br>2009 May 1 & 16<br>2009 Jun 16<br>2010 Feb 11-13<br>2010 Mar 9<br>2010 Apr 2 | Kong et al. 2006, Genzel et al. 2006; 2008; Cresci et al. 2009; Förster Schreiber et al. 2009 |
| ZC782941 (z=2.182) | K 0.05”x0.1” | NGS | 0.22” | 10h30<br>2007 Apr 16-23<br>2009 Apr 18<br>2010 Jan 9 & 13<br>2010 Feb 10 | Genzel et al. 2008; Cresci et al. 2009; Förster Schreiber et al. 2009; Mancini et al. in prep.; Peng et al. in prep. |
| ZC406690 (z=2.196) | K 0.05”x0.1”<br><br>K 0.125”x0.25” | NGS<br><br>Seeing limited | 0.22”<br><br>0.5“ | 10h00<br>2010 Apr 17<br>2010 May 25<br>2010 Nov 30<br>2010 Dec 7, 10, 29, 30, & 31,<br>2011 Jan 2 & 3<br><br>1h00; 2009 Dec 30 | Mancini et al. (in prep); Peng et al. (in prep.) |

## Table 2. Derived galaxy properties

| Source | | **BX599 all** | **BX482 clump A** | **D3a15504 clumps A-F** | **ZC782941 clump A** | **ZC406690 clump A** | **ZC406690 clump B** | **ZC406690 clump C** |
|---|---|---|---|---|---|---|---|---|
| z | 1 | 2.33 | 2.26 | 2.38 | 2.18 | 2.2 | 2.2 | 2.2 |
| $D_L$ (Gpc) | 2 | 19.1 | 18.3 | 19.6 | 17.6 | 17.7 | 17.7 | 17.7 |
| kpc/" | 3 | 8.33 | 8.38 | 8.3 | 8.42 | 8.41 | 8.41 | 8.41 |
| $F_{obs}$(Hα) 1e-16 erg/s/cm$^2$ | 4 | 3.3 | 0.35 | 0.04 | 0.2 | 1.4 | 0.57 | 0.4 |
| A(Hα)[1] | 5 | 0.73 | 1.1 | 1.8 | 2.1 | 1.1 | 1.1 | 1.1 |
| L(Hα)$_0$ erg/s[2] | 6 | 2.8e43 | 3.8e42 | 9.4e41 | 5.2e42 | 1.4e43 | 5.7e42 | 4.0e42 |
| SFR $M_\odot$/yr[3] | 7 | 66 | 12 | 3.3 | 17 | 40 | 11 | 14 |
| $M_{mol\text{-}gas}$ $M_\odot$ [3] | 8 | 3.3e10 | 7.8e9 | 3.0e9 | 8.7e9 | 1.6e10 | 7.8e9 | 9.6e09 |
| $\Sigma_{mol\text{-}gas}$ $M_\odot$pc$^{-2}$ [3] | 9 | 4.4e3 | 2.1e3 | 6.9e2 | 4e3 | 8.4e3 | 1.4e3 | 1.8e3 |
| $R_{HWHM\text{-}intr}$[3] kpc | 10 | 1.5 | 1 | 1 | 0.8 | 0.8 | 1.2 | 1.2 |
| $\Sigma_{star\text{-}form}$ $M_\odot$/yr/kpc$^2$ | 11 | 4.6 | 2.7 | 0.72 | 5.7 | 13.6 | 1.6 | 2.2 |
| $f_{broad}$ | 12 | 0.5(0.13) | 0.32(0.08) | 0.26(0.15) | 0.31(0.1) | 0.4(0.1) | 0.6(0.1) | ≤0.25 |
| [SII] 6718/6733 | 13 | - | 0.7(0.2) | 0.9(0.3) | 1.1(0.35) | 0.75(0.07) | 1.09(0.1) | - |
| $\sigma_{clump}$ km/s[4] | 14 | 76(20) | 62(3.4) | 53(7) | 95(7) | 81(4) | 88(4) | 78(3) |
| n(e)$_{clump}$[5] cm$^{-3}$ | 15 | - | 2000(+∞, -1000) | 900(+2500, -700) | 400(+1100, -350) | 1500(+900, -400) | 420(+230, -140) | - |
| $\gamma_{red}$[6] | 16 | 1 | 1.5 | 1.5 | 2 | 2 | 2 | 2 |
| L(Hα)$_{broad,0}$ erg/s | 17 | 1.4e43 | 1.8e42 | 3.7e41 | 3.2e42 | 1.1e43 | 6.9e42 | ≤2.0e42 |
| $\Delta v_{max}$ km/s | 18 | 1000 | 350 | ~400 | 420 | 440 | 810 | - |
| $M_{broad}$ $M_\odot$ | 19 | 4.5e8 | 6e7 | 1.2e7 | 1.1e8 | 3.6e8 | 2.2e8 | ≤6.5e7 |
| $dM_{out}/dt$ (case 1)[7] $M_\odot$/yr | 20 | 300 | 21 | 6 | 54 | 200 | 150 | ≤22 |
| $dM_{out}/dt$ (case 2)[7] $M_\odot$/yr | 21 | 68 | 5 | 1.4 | 13 | 46 | 34 | ≤5 |
| $dM_{out}/dt$ (case 3)[7] $M_\odot$/yr | 22 | 94 | 6.5 | 2.2 | 17.5 | 62 | 49 | ≤7 |
| $dM_{out\ 1/2}/dt$ / SFR [8] | 23 | 2.8 | 1.0 | 1.1 | 2.0 | 3.1 | 8.4 | ≤0.9 |
| $t_{expulsion}$[9] Myr | 24 | 360 | 1.2e3 | 1.6e3 | 520 | 265 | 170 | <1.5e3 |
| $t_*$ Myr[10] | 25 | - | 30-100 | >1e3 | | 80-800 | 100-3e4 | 80-800 |
| $t_Z$(closed)[10] Myr | 26 | 360 | 360 | 930 | 350 | 150 | 560 | 400 |
| $t_Z$(leaky)[10] Myr | 27 | 920 | 480 | 1600 | 650 | 260 | 2e4 | 510 |
| $t_{expansion}$ [10] Myr | 28 | 120 | 310 | 360 | 140 | 86 | 120 | - |
| $t_{diss}/t_{orbit}$ | 29 | 10 | 12 | 14 | 7 | 1.7 | 1.1 | ≥10 |
| $M_{*,final}/M_{gas,0}$ [11] | 30 | 0.27 | 0.49 | 0.48 | 0.34 | 0.25 | 0.11 | ≥0.52 |
| Δv/(sini 2 $R_{clump}$) [12] km/s/kpc | 31 | - | 19(-10) | 30(±12) | 42(10) | 20(-30) | 60(+30) | 30(+15) |
| 4.4 $M_{dyn\text{-}rot}$[13] $M_\odot$ | 32 | - | 4.3e8 | 1.1e9 | 1.1e9 | 2.2e8 | 2.1e9 | 2.1e9 |
| $M_{mol\text{-}gas}$/ 4.4 $M_{dyn\text{-}rot}$ | 33 | - | 18 | 3 | 8 | 75 | 4 | 5 |
| $M_{dyn\text{-}press}$ [11] $M_\odot$ | 34 | - | 2.1e9 | 1.5e9 | 3.9e9 | 2.8e9 | 5.2e9 | 3.9e9 |
| $M_{mol\text{-}gas}$/ ($M_{dyn\text{-}rot}$ + $M_{dyn\text{-}press}$) | 35 | - | 3.6 | 1.7 | 2.1 | 5.7 | 1.4 | 2.2 |
| $F_{rad}$= L/c dynes | 36 | 8.5e34 | 1.6e34 | 4.2e33 | 2.2e34 | 5.1e34 | 1.4e34 | 1.8e34 |
| ($\Delta v_{max}$ $dM_{out\ 1/2}/dt$)/ $F_{rad}$ | 37 | 14 | 2 | 3 | 4 | 7 | 34 | 2 |

**Footnotes for Table 2**

[1] $A(H\alpha)=7.4\ E(B-V)$, with $E(B-V)_{stars}=0.44\ E(B-V)_{gas}$ (Calzetti 2001)

[2] extinction corrected

[3] SFR ($M_\odot$/yr) $=L(H\alpha)_0/(2.1e41$ erg/s), $M_{mol-gas}(M_\odot)=1.2e9\ SFR(M_\odot/yr)^{0.75}\ R(kpc)^{0.54}$ (equation 2, Kennicutt et al. 2007). $L(H\alpha)_0$ is extinction corrected. Radii here and elsewhere in the table (e.g. row 10) are 'intrinsic' radii, with the instrumental resolution removed in squares

[4] intrinsic local velocity dispersion, after removal of beam smeared rotation and instrumental resolution

[5] from [SII] 6718/6733 ratio (Osterbrock 1989)

[6] correction for intrinsic differential extinction

[6] $\Delta v_{max}=\langle v_{broad}\rangle - 2\ \sigma_{broad}$

[7] estimates of outflow rates (Appendix B) for two models of photodissociation/case B recombination and for collisional excitation

[8] uses the average of the estimated of the two photodissociation/case B models in rows 20 and 21

[9] time scale for expulsion of gas by outflows: $t_{expulsion}= 2\ M_{mol-gas}/\ (dM_{out\ 1/2}/dt)$. The lifetime of a clump is shorter, given that in addition to gas outflows there is also star formation

[10] time estimates from stellar age dating (4.2.2), chemical enrichment (4.2.3, Appendix C) and expansion (4.2.4)

[11] ratio of final (stellar) mass at the time when all the gas is expelled by winds, relative to the initial gas mass, $M_{*,final}/M_{gas,\ t=0}=1/(1+[(dM_{out}/dt)/SFR])$

[12] maximum observed velocity gradient across clump in 'raw' velocity maps (in parentheses 'residual' maps); positive sign is prograde and negative sign retrograde with galaxy rotation

[13] $M_{dyn\text{-}rot}(M_\odot)=b\ 2.31e5\ (R_{HWHM}(kpc))^3\ (\Delta v\ (km/s)/(2\ \sin i\ R_{HWHM}(kpc)))^2$

[14] $M_{dyn\text{-}press}(M_\odot)=b\ 5.63e5\ (\sigma_{clump}(km/s))^2\ R_{HWHM}(kpc)$

# Table 3. Abundance measurements

| SOURCE | [NII]/Ha | Δ(NII/Ha) | μ=12 + log (O/H)[1] | Δμ |
|---|---|---|---|---|
| 1 | 2 | 3 | 4 | 5 |
| BX599 all | 0.19 | 0.08 | 8.49 | 0.18 |
| BX482 clump A | 0.14 | 0.017 | 8.41 | 0.05 |
| BX482 clumps B+C | 0.11 | 0.024 | 8.35 | 0.09 |
| BX482 nucleus | 0.22 | 0.027 | 8.53 | 0.05 |
| D3a15504 clumps A-F | 0.31 | 0.02 | 8.61 | 0.03 |
| D3a15504 interclump | 0.33 | 0.02 | 8.63 | 0.03 |
| D3a15504 nucleus | 0.43 | 0.04 | [8.69][2] | 0.04 |
| ZC782941 clumpA | 0.18 | 0.026 | 8.48 | 0.06 |
| ZC782941 clumps B-E | 0.28 | 0.021 | 8.58 | 0.03 |
| ZC782941 interclump | 0.205 | 0.021 | 8.51 | 0.04 |
| ZC406690 all | 0.097 | 0.017 | 8.32 | 0.08 |
| ZC406690 clumpA | 0.073 | 0.015 | 8.25 | 0.09 |
| ZC406690 clumpB | 0.22 | 0.017 | 8.53 | 0.03 |
| ZC406690 clumpC | 0.14 | 0.019 | 8.41 | 0.06 |

[1] $\mu=8.90+0.57 \log([NII]/H\alpha)$ (Pettini & Pagel 2004), with $\mu_\odot=8.66$ (Asplund et al. 2004)
[2] suspect because of possible influence of central AGN (Genzel et al. 2006)